\documentclass[12pt]{article}
\usepackage[margin=1.25in]{geometry} 
\usepackage{amsmath,amsthm,amssymb,todonotes}
\usepackage{multirow}
\usepackage{times}
\usepackage{bm}
\usepackage{listings}
\usepackage{graphicx}
\usepackage{subcaption}
\usepackage{enumerate}
\usepackage[
    colorlinks=true,
    linkcolor=blue,
    citecolor=blue,
    urlcolor=blue
]{hyperref}
\usepackage{dsfont}
\usepackage{float}
\usepackage[title]{appendix}
\usepackage{booktabs}
\usepackage{caption}
\usepackage{tikz}
\usetikzlibrary{decorations.pathreplacing}
\usetikzlibrary{arrows,shapes}
\usetikzlibrary{arrows.meta,bending,positioning}

\usepackage{pifont}
\usepackage{xcolor}

\usepackage{chngcntr}
\usepackage{leftindex}

\usepackage{natbib}

\usepackage[title]{appendix}

\DeclareMathOperator*{\argmax}{arg\,max}

\usepackage{bbm}

\usepackage{setspace}
\theoremstyle{plain}

\newtheorem{proposition}{Proposition}
\newtheorem{theorem}{Theorem}
\newtheorem{corollary}{Corollary}

\newtheorem{lemma}{Lemma}

\newtheorem{assumption}{Assumption}
\newtheorem{definition}{Definition}

\theoremstyle{remark}

\makeatletter

\newcommand{\Rmnum}[1]{\expandafter\@slowromancap\romannumeral #1@}

\title{Dynamic Reward Design\footnote{I am deeply grateful to Jeffrey Ely, Alessandro Pavan,
Piotr Dworczak, and Yingni Guo for their constant guidance and
support, as well as to Ian Ball, Xiaoyu Cheng, Maciej
Kotowski and Bruno Strulovici
for many detailed comments and discussions. For helpful comments, I
thank Nageeb Ali, Meghan Busse, H\'ector Chade, Eddie Dekel, Andrew
Ferdowsian, Tan Gan, Kyungmin (Teddy) Kim, Peter Klibanoff, Marciano Siniscalchi, Bal\'azs Szentes, Udayan Vaidya, Maren Vairo, and Asher Wolinsky. I also thank seminar and conference participants at Northwestern University, Tsinghua University,
Western University, University of Notre Dame, UC Riverside, University of
Hong Kong, Shanghai University of Finance and Economics, Indiana
University Kelley School of Business, Florida State University,
35th Stony Brook International Conference on Game Theory, 2025
Kansas Workshop in Economic Theory, 2025 SITE Conference,
2025 UNC Theory Conference, 2025 Midwest Trade and Theory
Conference, 2025 ASU Theory Conference, and 2026 North
American Summer Meeting of the Econometric Society.}}

\author{Yijun Liu\footnote{Department of Economics, University of Notre Dame, yliu2346@nd.edu.}}
\date{\today}

\begin{document}
\maketitle

\begin{abstract}
    We study a dynamic mechanism design problem with limited liability. A principal hires an agent to work over a finite horizon. The agent's costs are i.i.d.\ across periods and privately known, while his working status is publicly observable. The nonnegative-payment constraint distinguishes our problem from the dynamic screening literature. We identify conditions under which the optimal mechanism takes a simple form: a consecutive-working menu, in which the agent chooses when to start working and is then incentivized to work consecutively until the end. We explain why this consecutive structure is optimal despite i.i.d.\ costs and additively separable payoffs, shedding light on real-world reward schemes that require consecutive actions.
\end{abstract}

\newpage

\section{Introduction} \label{sec:introduction}

Consider the following scenario. Many firms today are concerned about their employees working remotely. Suppose a firm wants to incentivize a worker to come to the office using monetary rewards. The worker's cost of coming to the office is his private information; it is stochastic and may vary from day to day. The firm's objective is to maximize the number of days the worker comes to the office, net of the cost of rewards. What is the optimal reward scheme? Should the firm reward the worker based only on the total number of days he comes to the office? Should the rewards also depend on which days he comes? Or should the firm perhaps bundle certain days together?

This simple example captures the central question studied in this paper: how should a principal incentivize an agent to take observable actions when the agent faces private and stochastic costs and payments must be nonnegative? The framework applies to a broad range of settings, including the design of labor contracts and reward programs. For example, ride-sharing platforms such as Uber and Lyft face this problem when designing bonuses to incentivize drivers to complete more rides: the platform can observe whether a driver accepts or rejects a ride, but not the driver's cost of providing it. What, then, should the optimal reward scheme look like? Formally, this is a dynamic mechanism design problem. What distinguishes it from the existing literature is the nonnegative payment requirement, which we refer to as the \emph{limited liability} constraint. It is for this reason that we call the problem \emph{dynamic reward design}: a ``reward,'' by definition, is nonnegative.

In this paper, we identify conditions under which the optimal reward scheme takes a simple form: a \emph{consecutive-working menu}. Under this menu, the agent chooses when to begin working, and once he starts, he is incentivized to work consecutively until the final period.\footnote{The final period can be interpreted as the end of a bonus program.} It is equivalent to a menu of starting periods, since the finishing period is always the last. The resulting payment is easy to implement, as it depends on a single dimension of information: the number of consecutive periods the agent has worked, or equivalently, his starting period. By contrast, the most general payment scheme can be far more complicated, depending not only on how many periods the agent has worked but also on which periods he has worked.

Similar reward schemes appear in many real-world settings. For instance, Uber offers bonuses for completing a required number of consecutive trips. Lyft similarly rewards drivers for ride streaks. Starbucks awards consumers ``stars'' based on the number of consecutive days they make a purchase. Microsoft's Xbox rewards video game players based on the number of consecutive weeks they log in and complete tasks. Duolingo rewards users for completing lessons on consecutive days.\footnote{The Online Appendix provides further details on these examples.}

The puzzle is: what is the benefit of requiring consecutive actions? In our model, payoffs are additively separable across periods and the agent's costs are assumed to be i.i.d., so there is no direct complementarity across periods that would favor consecutiveness. What, then, is the economic intuition behind it?

The first contribution of this paper is to solve for the optimal mechanism and provide intuition for the consecutive structure. The second contribution is methodological. Limited liability makes the design problem nonstandard, and existing mechanism design techniques do not apply. We develop a novel approach to solve this problem, which may be useful for other design problems with limited liability.

\subsection{Discussion of intuition} \label{sec:introdiscussion}

First, the principal's ability to charge the agent is a powerful tool in dynamic mechanism design. To see this, consider the example illustrated in Figure \ref{fig:deposit}. Suppose the principal charges the agent a \$100 deposit on day 0 and promises to return it on day 2 if the agent follows the prescribed work instruction on day~1.  The instruction requires the agent to work for all cost realizations on day~1, given that his cost is drawn from the uniform distribution on $[0, 10]$. In return, the principal pays him \$5 for working. Since the expected cost is also \$5, a risk-neutral agent is willing to accept the contract ex ante. However, it is possible that the agent draws a high cost on day~1, say \$10. Then, he will still choose to work to avoid forfeiting the deposit. In this way, the principal implements the first-best outcome: the agent works with certainty, yet the principal pays only the average cost.

        \begin{figure}[h]
\centering
\begin{tikzpicture}[x=5.5cm,y=5.5cm]

    \draw[thick] (0,0) -- (1.2,0);
    \foreach \x in {0,0.6,1.2}
        \draw (\x,0) -- (\x,0.04);

    \node at (0,0.16) {Day 0};
    \node at (0,-0.12) {-\$100};

    \node at (0.6,0.16) {Day 1};
    \node at (0.6,-0.12) {work for sure};
    \node at (0.6,-0.22) {$\theta \sim u[0,10]$};
    \node at (0.6,-0.32) {pay: \$5};

    \node at (1.2,0.16) {Day 2};
    \node at (1.2,-0.12) {\$100};

\end{tikzpicture}
\caption{Example of a deposit.}
\label{fig:deposit}
\end{figure}

If deposits are allowed, the problem is trivial: the principal can simply achieve the first-best outcome in every period given the i.i.d.\ costs. Under the limited liability constraint, however, the principal loses this powerful tool. Limited liability not only complicates the design problem substantially but also reduces the principal's payoff. Yet there is still something the principal can do to discipline the agent, and it uses a similar logic. Specifically, the principal can backload some payments to the end; these payments effectively play the role of a deposit. These promised but postponed payments give the principal leverage: if the agent quits, he forfeits them. This leverage can be used repeatedly in subsequent periods to extract surplus. In fact, the promised payments build up over time, giving the principal increasing incentive power to regulate the agent.

Let us use a simple example to illustrate the intuition. Suppose the agent's cost of working is drawn from the uniform distribution on $[0,1]$, and the principal's value from work is $\alpha=1.2$. If they interact for only one period, it is optimal for the principal to post a wage of $0.6$. The agent then works if and only if his cost is below $0.6$, resulting in a $60\%$ probability of work. What happens if they interact over multiple periods? What additional strategies can the principal use to increase the working frequency while reducing the payment? Consider two periods. The wage payment from the first period can serve as a deposit held by the principal; it will be returned to the agent only if he continues working in the second period. For instance, the principal can commit to paying the agent his average cost, $0.5$, and the deposit of $0.6$ if the agent continues to work in the second period. However, if the agent shirks in the second period, the principal pays nothing and confiscates the deposit. This is equivalent to offering a wage of $1.1$ for working in the second period, which ensures that the agent works with certainty while minimizing the wage payment to $0.5$. The principal thus achieves the first best in the second period.

This simple example captures one part of the intuition for consecutive working: the principal can use compensation for past work to incentivize future effort. Specifically, under certain distributional assumptions, the promised payment for initiating work under the consecutive-working menu is large enough to incentivize the agent to work with certainty in the next period, while allowing the principal to achieve the first-best outcome. This amount is backloaded to the end and can be repeatedly used in every subsequent period to extract all of the agent's surplus in these periods.

The other part of the intuition is that assigning the agent to work consecutively after he starts does not affect the principal's expected payoff from other history paths. Specifically, we show that only the action rules at each starting period affect the agent's information rent across history paths; the action rules in subsequent periods do not. Combining these two parts, assigning the agent to work consecutively after he starts allows the principal to achieve the first-best outcome in the subsequent periods without affecting the expected payoff from other history paths. This second part of the intuition is less obvious, as it requires solving the dynamic optimization problem and characterizing the interdependence across different history paths. We provide a more detailed discussion in Sections~\ref{sec:mainresults} and~\ref{sec:proofmainresult}.

To see what a consecutive-working menu looks like, consider a simple three-period example. The agent's cost of working in each period is an i.i.d.\ draw from the uniform distribution on $[1,2]$, and the principal's value of work is $\alpha = 2$. Given the i.i.d.\ costs and additively separable payoffs, a natural benchmark is to repeat the static optimal mechanism: posting a wage of $1.5$ for working in each period. Under this naive mechanism, the principal's expected payoff is $0.75$. In contrast, the optimal mechanism yields an expected payoff of $1.01$. This optimal mechanism is a consecutive-working menu, illustrated in Figure~\ref{fig:consecutiveworkingmenu}.

\begin{figure}[ht]
\centering
\begin{tabular}{c c c c c}
\toprule
 Period 1 & Period 2 & Period 3 & Payment \\
\midrule
 \textcolor{green!60!black}{\ding{51}} & \textcolor{green!60!black}{\ding{51}} & \textcolor{green!60!black}{\ding{51}} & \textcolor{blue}{\$4.82} \\[6pt]
\textcolor{red}{\ding{55}} & \textcolor{green!60!black}{\ding{51}} & \textcolor{green!60!black}{\ding{51}} & \textcolor{blue}{\$2.76} \\[6pt]
 \textcolor{red}{\ding{55}} & \textcolor{red}{\ding{55}} & \textcolor{green!60!black}{\ding{51}} & \textcolor{blue}{\$1.14} \\[6pt]
 \multicolumn{3}{c}{All other working patterns} & \textcolor{blue}{\$0} \\
\bottomrule
\end{tabular}

\vspace{4pt}
{\small \textcolor{green!60!black}{\ding{51}} = work, \quad \textcolor{red}{\ding{55}} = shirk}

\caption{The optimal consecutive-working menu.}
\label{fig:consecutiveworkingmenu}
\end{figure}

The agent can choose to start working in any of the three periods, and once he starts, he is required to work consecutively until the end. Any other working pattern generates zero payment. For example, if the agent chooses (work, shirk, work), he receives nothing. More specifically, the menu shown above is the equivalent menu implementation of the optimal mechanism, which takes the form of cutoff rules based on the agent's working status history. If the agent starts working in the first period, the cutoff for the next period is the highest cost, $2$, so the mechanism prescribes the agent to work with certainty in the second period. Intuitively, the promised payment from the first period is already large enough to allow the principal to achieve the first-best outcome in the second period. If the agent instead shirks in period~2 after starting in period~1, he deviates to an off-equilibrium path, and the principal should pay him nothing.

This consecutive working requirement coincides with many real-world reward schemes. For example, Uber's driver bonus program requires drivers to ``complete the required number of trips without rejecting, unfulfilling, or going offline between trips. Your extra earnings will be added to the last trip receipt''.\footnote{See
\href{https://help.uber.com/en-GB/driving-and-delivering/article/consecutive-trips-issue?nodeId=13ca6726-fa40-40c5-96e4-14a2aed9ea06}
{Uber's description of its Consecutive Trips program}
for the exact terms and conditions.} One might argue that this requirement is intended for habit formation or to prevent drivers from switching to competing platforms. Our result shows that, on top of these good reasons, the principal's profit maximization itself can also lead to the consecutive working requirement.

\subsection{Overview of the main results}

Theorem~\ref{thm:optimalmechproperties} shows that under certain distributional assumptions (Assumption~\ref{assumption1}), the optimal contract takes one of two forms: a consecutive-working menu or an always-working mechanism. Both are menus of starting periods and feature consecutive working until the end. The difference is that the consecutive-working menu offers all $N$ periods as starting periods, while the always-working mechanism offers only the first period. Note that the consecutive-working menu distorts the starting time but not the actions after starting. The principal may optimally induce the agent to delay starting for many periods, despite the costs being i.i.d. In contrast, without the limited liability constraint, the agent is induced to start working immediately.

Next, Theorem~\ref{thm:mainresult} shows that the magnitude of $\alpha$, the principal's value of work, determines which form is optimal. The optimal mechanism is the always-working mechanism when $\alpha$ is above a threshold and a consecutive-working menu when $\alpha$ is below it.

Moreover, we characterize the general structure of the optimal mechanism without any distributional assumptions in Theorem~\ref{thm:generaloptimalmech}. It is a two-stage mechanism: in the first stage, the agent is assigned to work when the cost is low and allowed to shirk when the cost is high; in the second stage, the agent is assigned to work consecutively until the end. The consecutive-working feature is thus a robust property of the optimal mechanism. The first stage serves to accumulate promised payments; once they are large enough, the mechanism enters the second stage, in which the principal achieves the first-best outcome in every period. Only the payments for the first stage need to be backloaded to the end, while payments for the second stage can be made each period. We therefore interpret the first stage as a trial period and the second stage as a permanent employment period.

Finally, in the extension section, we show that stochastic mechanisms can improve upon the optimal deterministic mechanism when the latter is a consecutive-working menu. The idea of using past compensation to incentivize future effort still applies, but in a more flexible and powerful way. The principal can reallocate promised payments across different realizations of actions, thereby achieving better overall incentive power across scenarios.

The rest of the paper is organized as follows. Section~\ref{sec:relatedliterature} discusses the related literature. Section~\ref{sec:model} introduces the model, and Section~\ref{sec:dynamicopt} sets up the dynamic optimization problem. Section~\ref{sec:mainresults} presents and discusses the main results. Section~\ref{sec:proofmainresult} develops the solution method. Section~\ref{sec:properties} discusses properties of the optimal mechanism: comparative statics in Section~\ref{sec:comparativestatics}, the general optimal mechanism in Section~\ref{sec:generaloptimalmech}, and the role of backloading payments in Section~\ref{sec:backloading}. Finally, Section~\ref{sec:extensions} explores an extension to stochastic mechanisms, and Section~\ref{sec:conclusion} concludes. All proofs are in the Appendix, except those for Section \ref{sec:extensions}, which are
provided in the Online Appendix.

\subsection{Related literature} \label{sec:relatedliterature}

This paper contributes to the literature on dynamic mechanism design (see, e.g., \cite{BARON1984}, \cite{BESANKO198533}, \cite{CourtyandLi2000}, \cite{Battaglini2005}, \cite{Eső2007}, \cite{Boleslavsky2012}, \cite{pavan2014dynamic}, \cite{BergemannandVälimäki2019}). As in these studies, this paper seeks to find the revenue-maximizing mechanism given the agent's time-evolving private information. The key difference is that these papers consider persistent private information without limited liability, whereas we consider independent private information with limited liability.

This paper differs from the existing literature in two key respects. First, we impose a limited liability constraint. While this constraint is standard in labor contracts, it has not been widely studied in the dynamic mechanism design setting. Second, we consider independent types, whereas most existing work focuses on persistent types. The focus in that literature is on how the persistence of private types influences information rents (as captured by the impulse response function in \cite{pavan2014dynamic}) and the corresponding optimal mechanisms. In fact, with independent costs, the design problem is trivial without the limited liability constraint. However, the combination of independent costs and limited liability gives rise to a nontrivial design problem.

Our work is also related to the literature on dynamic mechanism design with nonnegative stage payoffs, including \cite{krishna2013stairway}, \cite{krasikovandlamba2021}, and \cite{krahmer2022dynamic}. Their constraint is stricter than ours: they require that the payment exceed the agent's cost in each period, whereas we require only that the payment itself be nonnegative. Our limited liability constraint thus occupies the middle ground between the nonnegative stage payoff constraint and no constraint on payments.

In addition, the limited liability constraint makes the participation constraint more difficult to handle. The literature typically needs to ensure only the ex-ante IR constraint, as the principal can charge the agent a large deposit upfront. In our setting, however, it is unclear where the participation constraints bind, so we must impose the interim IR constraints for every history. From this perspective, this paper is also related to work in dynamic mechanism design that pays special attention to IR constraints. For instance, \cite{BergemannandVälimäki2010} also account for interim IR constraints in their analysis, but do not impose limited liability. \cite{krähmerandstrausz2015} and \cite{BERGEMANN2020105055} impose ex-post IR constraints to address consumer withdrawal rights in dynamic selling contexts. \cite{mirrokni2020nonclairvoyant} and \cite{ashlagi2023} impose ex-post IR constraints in dynamic auction and multi-product selling contexts, respectively.

Limited liability is a natural assumption in many dynamic labor and financial contracting settings. A large literature studies optimal contracts under limited liability in these contexts, but most of it does so from the perspective of moral hazard (see, e.g., \cite{Clementi2006}, \cite{DeMarzoandSannikov2006}, \cite{Biais2007}, \cite{fong2017relational}). In contrast, we study the optimal labor contract with limited liability from the perspective of adverse selection. \cite{Board2011} examines the dynamically optimal contract for a holdup problem in which the agent has no private information. The optimal contract in that setting shares a key similarity with ours: the agent earns rent only in the first trade (or work), after which the allocation becomes efficient.\footnote{Since the agent does not possess private information, the rent in \cite{Board2011} is not an information rent; rather, it is compensation provided to prevent the agent from holding up the principal.}

This paper also relates to work on dynamic contracts in other settings, such as \cite{Halac2016} on experimentation and \cite{GershkovandMoldovanu2014} on revenue management. Finally, this paper connects to the literature on linking mechanisms, including \cite{Jacksonsonnenschein2007}, \cite{Frankel2014, FRANKEL2016}, and \cite{ball2024}.\footnote{While these studies require either a large or infinite number of periods to apply the law of large numbers or achieve stationarity, we focus on optimality for any finite number of periods.} In that literature, the optimal mechanism is a quota mechanism that links the agent's actions across periods, whereas our optimal mechanism features consecutive linking.

\section{Model} \label{sec:model}

A principal (she) hires an agent (he) to work in $N$ periods. Denote $\{1, \ldots, N\}$ as $\mathcal{N}$. The agent's cost of working is his private information, while his action of working or shirking is publicly observable. The agent's cost of working in each period, $\tilde{\theta}_t$, is an independent and identically distributed (i.i.d.) random variable with probability density function $f$ and cumulative distribution function $F$, supported on $\Theta\equiv[\underline{\theta},\bar{\theta}]$, where $0\leq\underline{\theta}<\bar{\theta}<\infty$.  We assume that $f$ is continuous and $f(\theta) >0$ for all $\theta \in \Theta$. We use a tilde to differentiate between random variables and their realizations; for example, $\tilde{\theta}_t$ represents the random variable and $\theta_t$ represents its realization. Let $h_t \equiv (\theta_1, \theta_2, \ldots, \theta_t) \in \Theta^t$ denote the history of realized costs up to period $t$.\footnote{We let $h_0$ denote $\emptyset$.} Let $\bm{\theta} \equiv (\tilde{\theta}_1, \tilde{\theta}_2, \ldots, \tilde{\theta}_N)$ denote the random vector of the agent's complete type profile. We use $E_{\pmb{\theta}}(\cdot)$ for the expectation over $\bm{\theta}$ and $E_{\pmb{\theta}}^t(\cdot)$ for the expectation over the future costs $  (\tilde{\theta}_{t+1}, \tilde{\theta}_{t+2}, \ldots, \tilde{\theta}_N)$ after $h_t$ is realized.

At the beginning of the game, the principal commits to a dynamic mechanism $\{x_t,p_t\}_{t=1}^N$. By the Revelation Principle, we restrict attention to direct mechanisms in which the agent reports truthfully and follows the recommended actions on the equilibrium path. In period $t$, the action rule $x_t:\Theta^t \rightarrow \{0,1\}$ maps the report history \(h_t\) to a recommended action, where $1$ denotes work and $0$ denotes shirk. We impose the limited liability constraint, which requires all payments to be nonnegative. Accordingly, the period-\(t\) payment rule $p_t:\Theta^t \rightarrow \mathbb{R}_{+}$ maps \(h_t\) to a nonnegative payment.\footnote{Neither the action rule nor the payment rule depends on the agent's historical actions for the following reasons. First, the obedience constraints ensure that the agent will follow the prescribed action in each period. Second, the past actions do not affect the agent's future type realizations. Thus, even if we formulate the mechanism as a function of the agent's historical actions, the past actions should be functions of the historical types themselves.} The principal makes the payment \(p_t(h_t)\) after the period-\(t\) action is publicly observed and only if the agent has followed the recommended action. Without loss of generality, we assume that if the agent deviates, the principal terminates the game and withholds all current and future payments.

The agent's expected payoff $U$ is defined as
\[
U = E_{\pmb{\theta}} \left[ \sum_{t=1}^N \left(p_t - \tilde{\theta}_t\cdot x_t \right) \right].
\]
The principal's expected payoff $V$ is defined as
\[
  V =  E_{\pmb{\theta}} \left[\sum_{t=1}^N \Big( \alpha \cdot x_t  - p_t \Big)\right].
\]
We assume $\alpha \geq \bar{\theta}$, making work always efficient.\footnote{Without this assumption, even the first-best allocation does not require consecutive working, so the optimal mechanism will not exhibit the consecutive-working feature.}

\subsection{Principal’s optimization problem} \label{sec:dynamicopt}

In this setting, a mechanism $\{x_t, p_t\}_{t=1}^N$ is implementable if it satisfies periodic incentive compatibility (IC-t), individual rationality (IR-t), obedience and limited liability (LL-t) constraints. Since we assume that the principal will end the game if the agent does not follow the recommendation, obedience is implied by the participation constraint. Let $u_t(h_{t-1},\theta_t, \hat{\theta}_t)$ denote the agent's expected payoff calculated in period $t$, assuming truthful future reports and given the past reports $h_{t-1}$, current cost realization $\theta_t$ and current report $\hat{\theta}_t$, i.e.,
\begin{align*}
    u_t(h_{t-1},\theta_t, \hat{\theta}_t) \equiv & \;  p_t(h_{t-1}, \hat{\theta}_t) - \theta_t \cdot x_t(h_{t-1},\hat{\theta}_t) +\\
    &E_{\pmb{\theta}}^t \left[ \sum_{i=t+1}^N p_i(h_{t-1},\hat{\theta}_t,\{\tilde{\theta}_k\}_{k=t+1}^i) - \tilde{\theta}_i \cdot x_i(h_{t-1},\hat{\theta}_t,\{\tilde{\theta}_k\}_{k=t+1}^i) \right].
\end{align*}
To simplify notation, we use $u_t(h_{t-1},\theta_t)$ to denote $u_t(h_{t-1},\theta_t, \theta_t)$. Definition \ref{def:implementability} formalizes the implementability conditions.

\begin{definition}\label{def:implementability}
    A mechanism $\{x_t, p_t\}_{t=1}^N$ is implementable if the following conditions hold:
    \[
    \theta_t \in \argmax_{\hat{\theta}_t \in \Theta} \; u_t(h_{t-1}, \theta_t,\hat{\theta}_t) \tag{IC-t}
    \]
    \[
      u_t(h_{t-1},\theta_t) \geq 0 \tag{IR-t}
    \]
    \[
    p_t(h_{t-1}, \theta_t) \geq 0 \tag{LL-t}
    \]
    for all $t$, $\theta_t \in \Theta$ and $h_{t-1}\in \Theta^{t-1}$.
\end{definition}

The dynamic mechanism design literature typically requires the agent to be on-path truthful, meaning it is optimal for the agent to report truthfully given truthful past reports. In our i.i.d.\,type setting, on-path truthfulness implies off-path truthfulness, indicating that it remains optimal for the agent to be truthful even if he has lied in the past. As discussed in \cite{pavan2014dynamic}, this is because the agent's expected payoff in any period $t$ depends only on his current true type and past reports, but not on his past true types. Consequently, we can impose a stronger IC constraint that requires the agent to be truthful across all histories.

In the literature without limited liability, it is often assumed that the principal can require the agent to post a sufficiently large bond in the initial period, making it unprofitable for the agent to quit and forfeit the bond mid-game. Consequently, the mechanism designer only needs to ensure the IR constraint is satisfied in period 1.\footnote{Or in period 0 before the realization of any private information, depending on the timing of the contract.} However, in our setting, limited liability prevents the principal from charging the agent a deposit. Therefore, the principal must ensure that the agent will not quit in any period by making the periodic IR constraints (IR-t) hold for all $t$.

The principal’s optimization problem is to find a dynamic mechanism $\{x_t, p_t\}_{t=1}^N$ that maximizes her payoff $V$, subject to the IC-t, IR-t, and LL-t constraints for all $t$. Formally, the principal’s problem can be written as follows:
\[
\max_{\{x_t, p_t\}_{t=1}^N} E_{\pmb{\theta}} \left[ \sum_{t=1}^N \alpha \cdot x_t - p_t \right]
\]
\[
\text{s.t.} \quad \text{IC-t, IR-t, LL-t} \quad \forall \; t.
\]

\section{Main Result} \label{sec:mainresults}

We now define the two classes of mechanisms that characterize our main result, and then state the main theorem. First let us use Figure \ref{fig:consecworkingexample} below to illustrate what a consecutive-working menu looks like in a three-period case, then we give the formal definition.

   \begin{figure}[h]
                    \centering
                    \begin{tikzpicture}[x= 5.5cm, y = 5.5cm]
                        \node at (0.6, 0.73) {$t = 3$};
                        \node at (0.3, 0.73) {$t = 2$};
                        \node at (0, 0.73) {$t = 1$};
            
                        \draw[thick] (0, -0.2) -- (0, 0.2);
                        \draw (-0.02,0.2) -- (0.02, 0.2);
                        \draw (-0.02, -0.2) -- (0.02, -0.2);
                        \node[above] at (0, 0.22) {$\overline{\theta}$}; 
                        \node[below] at (0, -0.22) {$\underline{\theta}$}; 
                        \node[left, red] at (0, 0.1) {shirk}; 
                        \node[left, blue] at (0, -0.1) {work}; 
                        \filldraw[black] (0, 0) circle (1.5pt) node[anchor= east]{$c_{1}$};
                
                        \draw [-Stealth, red] (0.05, 0.05) to (0.2, 0.22);
                        \draw [-Stealth, blue] (0.05, -0.1) to (0.25, -0.2);
                
                        \draw[thick] (0.3, 0.1) -- (0.3, 0.42); 
                        \draw (0.28, 0.42) -- (0.32, 0.42);
                        \draw (0.28, 0.1) -- (0.32, 0.1);
                        \filldraw[black] (0.3, 0.235) circle (1.5pt) node[anchor= west]{$c_{2}(0)$};
                
                        \draw[thick] (0.3, -0.1) -- (0.3, -0.4); 
                        \draw (0.28, -0.4) -- (0.32, -0.4);
                        \draw (0.28, -0.1) -- (0.32, -0.1);
                        \filldraw[black] (0.3, -0.1) circle (1.5pt) node[anchor= west]{$c_{2}(1)$};
                        
                        \draw [-Stealth, red] (0.35, 0.31) to (0.55, 0.52);
                        \draw [-Stealth, blue] (0.35, 0.15) to (0.55, 0.15);
                
                        \draw[thick] (0.62, 0.4) -- (0.62, 0.65); 
                        \draw (0.6, 0.4) -- (0.64, 0.4);
                        \draw (0.6, 0.65) -- (0.64, 0.65);
                        \filldraw[black] (0.62, 0.45) circle (1.5pt) node[anchor= west]{$c_{3} (0,0)$};
                
                        \draw[thick] (0.62, 0.27) -- (0.62, 0.0); 
                        \draw (0.6, 0.27) -- (0.64, 0.27);
                        \draw (0.6, 0.0) -- (0.64, 0.0);
                        \filldraw[black] (0.62, 0.27) circle (1.5pt) node[anchor= west]{$c_{3}(0,1)$};
                
                        \draw [-Stealth, blue] (0.35, -0.25) to (0.55, -0.25);
                
                        \draw[thick] (0.62, -0.1) -- (0.62, -0.4); 
                        \draw (0.6, -0.1) -- (0.64, -0.1);
                        \draw (0.6, -0.4) -- (0.64, -0.4);
                        \filldraw[black] (0.62, -0.1) circle (1.5pt) node[anchor= west]{$c_{3}(1,1)$};

                        \node at (0.6, 0.73) {$t = 3$};
            \node at (0.3, 0.73) {$t = 2$};
            \node at (0, 0.73) {$t = 1$};

                        \end{tikzpicture} 
                        \caption{The consecutive-working menu for three periods.}
                        \label{fig:consecworkingexample}
                \end{figure}

A consecutive-working menu is a cutoff mechanism. The agent is assigned to work if his cost is below a certain cutoff and to shirk otherwise. The cutoff $c_t$ in period $t$ depends on the past working status history. Here $1$ represents working and $0$ represents shirking. It has two defining features. First, the agent can begin working in any period with positive probability, as $c_1$, $c_2(0)$, and $c_3(0,0)$ are all interior points. Second, once the agent starts working, he is induced to work consecutively to the end, as all other cutoffs are equal to $\bar{\theta}$. It can be viewed as a menu of starting periods: the agent can choose the starting period from $\{1,2,3\}$, and then he is prescribed to work consecutively until the end.

The formal definition is given in Definition \ref{def:consecworking} below. Let $w_t$ denote the working status history up to period $t$, i.e., $w_t \equiv (x_1, x_2, \ldots, x_t) \in \{0,1\}^t$. Let $\mathbf{0}_t$ and $\mathbf{1}_t$ denote the length-$t$ vectors of zeros and ones, respectively. We write $c_t(\mathbf{0})$ as shorthand for $c_t(\mathbf{0}_{t-1})$, the cutoff for initiating work in period~$t$ given the agent has never worked before. We refer to $\{c_t(\mathbf{0})\}_{t=1}^{N}$ as the \emph{initiation cutoffs}.

\begin{definition}\label{def:consecworking}
    A consecutive-working menu is a cutoff mechanism such that, in each period $t$, the agent is assigned to work if $\theta_t \leq c_t(w_{t-1})$ and to shirk otherwise. The cutoffs satisfy:
    \begin{enumerate}[(i)]
        \item $c_t(\mathbf{0}) \in (\underline{\theta},\bar{\theta})$ for all $t \in \mathcal{N}$ (possible to start working in any period),
        \item $c_t(w_{t-1})=\bar{\theta}$ for all on-path  $w_{t-1} \notin \{\mathbf{0}_{t-1}, \emptyset\}$ (consecutive working after the start).
    \end{enumerate}
    There are no interim payments: $p_k = 0$ for all $k < N$. The final payment for work history $(\mathbf{0}_t,\mathbf{1}_{N-t})$ is:
    \begin{align}\label{eq:finalpayment}
        p_N = c_{t+1}(\mathbf{0})+\sum_{i=t+2}^{N}\int_{\underline{\theta}}^{c_i(\mathbf{0})}F(\theta)d\theta+(N-t-1)E(\theta)
    \end{align}
    for all $t<N-1$, $p_N=c_{N}(\mathbf{0})$ for $t=N-1$, and $p_N=0$ for $t=N$.
\end{definition}

The consecutive-working menu consists of $N$ starting options, $\{1, 2, \ldots, N\}$, where each option requires the agent to work consecutively from the chosen starting period until the end. The agent starts in period $t$ if his period-$t$ cost is below the cutoff $c_t(\mathbf{0})$ and his costs in all previous periods were above the respective cutoffs. The final payment depends on the starting period, or equivalently, the number of consecutive working periods.

\begin{definition} \label{def:alwaysworking}
    An always-working mechanism requires the agent to work for all $\theta_t \in \Theta$ and for all $t\in \mathcal{N}$. There are no interim payments. The final payment is $p_N=\bar{\theta}+(N-1) E(\theta).$
\end{definition}

\begin{assumption} \label{assumption1}
    $\underline{\theta} + E(\theta) \geq \bar{\theta}$.
\end{assumption}

Assumption~\ref{assumption1} requires that the width of the cost support not exceed the mean cost, i.e., $\bar{\theta} - \underline{\theta} \leq \mathbb{E}(\theta)$. This is satisfied when $\underline{\theta}$ is not too small relative to $\bar{\theta}$. For example, for the uniform distribution, the assumption is equivalent to $\underline{\theta} \geq \bar{\theta}/3$. Hence, it is satisfied by the uniform distribution on $[1,2]$ used in the introductory example. More generally, Assumption~\ref{assumption1} is implied by the stronger condition $\underline{\theta} \geq \bar{\theta}/2$, which requires that the cost distribution is not excessively dispersed, in the sense that the lower bound of the support is at least half of the upper bound.

\begin{theorem} \label{thm:optimalmechproperties}
     Under Assumption~\ref{assumption1}, the optimal mechanism is either a consecutive-working menu or an always-working mechanism.
    \end{theorem}

Theorem~\ref{thm:optimalmechproperties} shows that, under Assumption \ref{assumption1}, the optimal mechanism takes one of two simple forms. The always-working mechanism is a special case of the consecutive-working menu in which the only available starting period is the first period. The theorem shows that this is the only degenerate case that can be optimal. In other words, the optimal mechanism always prescribes consecutive work until the end; the set of feasible starting periods is either the full set \(\{1,2,\ldots,N\}\) or the singleton set \(\{1\}\). Theorem \ref{thm:mainresult} in Section \ref{sec:comparativestatics} further shows that, under Assumption \ref{assumption1}, there exists a threshold value \(\hat{\alpha}\) such that the always-working mechanism is optimal for \(\alpha \geq \hat{\alpha}\), and the consecutive-working menu is optimal for \(\alpha < \hat{\alpha}\).

The consecutive-working menu distorts the starting time, but not the actions thereafter. Specifically, because the cutoff for starting work is an interior point in every period, the agent begins working in each period with positive probability. As a result, despite the i.i.d.\ costs, the principal may optimally induce the agent to delay starting until any period, including the last, or never start working at all. In contrast, the optimal mechanism without limited liability induces the agent to start working immediately.

The intuition behind the optimal mechanism is twofold. First, under Assumption~\ref{assumption1}, the promised payment for initiating work is large enough for the principal to implement the first-best allocation in all subsequent periods once the agent starts working. Specifically, the payment in \eqref{eq:finalpayment} can be interpreted as follows. To induce the agent to start working in period~$t+1$ given no prior work, the principal promises a payment equal to the cutoff cost $c_{t+1}(\mathbf{0})$ plus an information rent of
$\sum_{i=t+2}^{N}\int_{\underline{\theta}}^{c_i(\mathbf{0})}F(\theta)\,d\theta$.
Thereafter, the principal compensates the agent only for the expected cost of working, $E(\theta)$, in each remaining period. Under Assumption~\ref{assumption1}, the payment for initiating work is large enough to induce the agent to work with certainty in the following period, and therefore in all subsequent periods. Moreover, this amount is backloaded to the end, serving to discipline the agent: if the agent shirks in any future period, he forfeits the payment. This leverage can be used repeatedly in every subsequent period, allowing the principal to extract the agent's entire surplus in all remaining periods.

Second, the less obvious part of the intuition is that the actions after the agent starts working have no interdependence effect on the information rent along other history paths. Let $u_1^*(\bar{\theta})$ denote the agent's first-period expected payoff under the optimal mechanism when he has the highest cost $\bar{\theta}$. It also serves as the constant term in the agent's expected payoff $u_t(h_t)$ after applying the envelope formula. 
Unlike in static mechanism design or dynamic settings without limited liability, $u_1^*(\bar{\theta})$ cannot simply be normalized to $0$; determining it is an important and nontrivial part of the analysis. We show that only the initiation cutoffs enter \(u_1^*(\bar{\theta})\) and therefore affect the agent's information rent for every history path; the cutoffs after the agent starts working do not. Consequently, assigning the agent to work consecutively after he starts allows the principal to achieve the first best in all subsequent periods without affecting any other history path, and is therefore optimal. A more technical explanation of these two parts of the intuition is provided in Section~\ref{sec:simplifyoptim}.

Assumption~\ref{assumption1} is sufficient but not necessary; it guarantees the feasibility of every consecutive-working menu. However, the conclusion of Theorem~\ref{thm:optimalmechproperties} requires only the feasibility of the optimal consecutive-working menu.\footnote{We show that whenever an optimal consecutive-working menu exists and is implementable, it is also optimal among all feasible mechanisms.} In Section~\ref{sec:simplifyoptim}, we derive a weaker condition that guarantees precisely this.\footnote{To see this, consider the two-period
example in Section \ref{sec:introdiscussion} (costs uniform on $[0,1]$, $\alpha=1.2$).
Assumption~\ref{assumption1} fails, yet the
weaker condition holds and the optimal mechanism is a
consecutive-working menu.} Because this condition depends on the endogenous optimal cutoffs,\footnote{We will discuss how to calculate them in Section \ref{sec:simplifyoptim}.} Assumption~\ref{assumption1} is stronger but simpler to verify. 
Indeed, starting directly from Assumption~\ref{assumption1} allows us to establish the consecutive-working pattern without fully solving the optimization problem. Solving the problem, however, is necessary to characterize the optimal mechanism more precisely, including the feasible starting periods, the payment rule, the optimal cutoffs, and their dependence on past histories. It also enables us to conduct comparative statics in the principal's value of work (Theorem~\ref{thm:mainresult}). Moreover, Theorem~\ref{thm:generaloptimalmech} shows that, without any distributional assumptions, the optimal mechanism has a two-stage structure with a consecutive-working second stage. Thus, consecutive working is a robust feature of the optimal mechanism.

\section{Solving the Dynamic Design Problem}\label{sec:proofmainresult}

We first provide an overview of the strategy. The limited liability constraint makes the design problem nontrivial and the standard methods cannot be directly applied. Specifically, it is unclear and nonstandard where the IR constraints bind. Therefore, we cannot simply rewrite the principal's problem as the maximization of virtual surplus, where one would typically maximize the integrand pointwise or apply ironing when the integrand is non-monotone. Instead, we must characterize the action rule for every history. Since there are infinitely many histories, this requires optimizing over infinitely many action rules (functions). To solve this problem, we proceed in three simplification steps.

First, in Section \ref{sec:thresholds}, we show that the problem can be reduced to optimizing over cutoffs. Given the binary action space, it is straightforward that the action rules take a cutoff form. The nontrivial part is characterizing the final payment \(p_N\). Unlike in standard screening problems, \(p_N\) cannot be determined from the action rules through a direct application of the envelope formula. Nevertheless, we show that it is still uniquely pinned down by the action rules through a more intricate characterization involving an infimum function.

Second, in Section \ref{sec:reducedim}, we reduce the dimensionality of the problem. We show that the agent's working status history is a sufficient statistic for the cost history in determining future cutoffs. As a result, the number of relevant cutoffs is reduced from infinitely many to finitely many. Moreover, the infimum in \(p_N\) becomes a minimum, transforming the original problem into a nested optimization problem consisting of an outer maximization and an inner minimization.

Third, in Section \ref{sec:simplifyoptim}, we solve the inner minimization problem by identifying the history whose IR constraint must bind, thereby transforming the problem into a standard constrained optimization problem. We then solve the relaxed problem obtained by dropping the constraints, whose solution is a consecutive-working menu. Finally, we identify conditions under which the relaxed solution is feasible and therefore also solves the original constrained problem. The detailed proofs are provided below.

\subsection{Optimization over cutoffs} \label{sec:thresholds}

\begin{lemma} \label{lem:ICconditions}
    A mechanism is incentive-compatible if and only if, for all $t$ and $h_{t-1}$,
    \begin{enumerate}[(i)]
        \item $x_t(h_{t-1}, \theta_t)$ is non-increasing in $\theta_t$,
        \item $u_t(h_{t-1},\theta_t)=u_t(h_{t-1},\bar{\theta})+\int_{\theta_t}^{\bar{\theta}}x_t(h_{t-1}, s)ds$. 
    \end{enumerate}
\end{lemma}

\begin{corollary} \label{cor:threshold}
    For all $t$ and $h_{t-1}$, the action rule $x_t(h_{t-1}, \theta_t)$ is a cutoff rule: $x_t(h_{t-1}, \theta_t)=1$ when $\theta_t \leq c_t(h_{t-1})$, and $x_t(h_{t-1}, \theta_t)=0$ when $\theta_t > c_t(h_{t-1})$.
\end{corollary}

Lemma \ref{lem:ICconditions} is a standard result derived from the envelope theorem in \cite{milgrom2002envelope}. Corollary \ref{cor:threshold} follows from the first condition in Lemma \ref{lem:ICconditions} and the fact that the action space is binary. To show that the problem can be transformed into an optimization problem over cutoffs, we will first show that it is without loss of generality to pay the agent only at the very end. Though the optimal mechanism does not necessarily require backloading all payments to the end, as discussed in Section \ref{sec:backloading}, having only one payment at the end simplifies the analysis.

\begin{proposition}\label{prop:payatend}
    It is without loss of generality to pay the agent only at the very end, i.e. $p_t = 0$ for all $t < N$.
\end{proposition}

The reason behind Proposition~\ref{prop:payatend} is that, for any implementable mechanism that pays the agent in interim periods, the principal can backload all payments to the last period while preserving implementability. This is because, first, backloading does not change the agent's incentive to report truthfully in each period. The total current and expected future payments for each report remain the same; they are merely postponed. More precisely, the expected payoff difference between truth-telling and lying is unchanged. Second, backloading makes the agent's participation constraints easier to satisfy, as the expected payoff from participating increases: payments that would have been made in earlier periods are deferred to the end, raising the agent's continuation payoff at every interim history.

Backloading payments is a common result in dynamic contracting with quasi-linear payoffs, for example, in other dynamic screening models such as \cite{krishna2013stairway}, \cite{ashlagi2023}, and \cite{krahmer2022dynamic}, as well as in dynamic moral hazard models with limited liability constraints, for example, \cite{Clementi2006}, \cite{DeMarzoandSannikov2006}, and \cite{Biais2007}. Here, backloading part of the payments is a necessary feature of the optimal mechanism, as it allows the principal to extract surplus from the agent by leveraging the promised payments. We discuss this intuition in the introduction of this paper. We will further elaborate on how much of the payments are required to be backloaded in Section \ref{sec:backloading}.

Proposition \ref{prop:payatend} allows us to simplify the principal's optimization problem by focusing on pay-at-the-end mechanisms, denoted as $(\{x_t\}_{t=1}^N, p_N)$, where $p_N: \Theta^N \rightarrow \mathbb{R}_{+}$.  Then, the principal's optimization problem can be rewritten as follows:
\[
\max_{\{x_t\}_{t=1}^N, p_N} E_{\pmb{\theta}} \left[ \sum_{t=1}^N \alpha \cdot x_t - p_N \right]
\]
\[
\text{s.t.} \quad \text{IC-t, IR-t} \quad \forall \; t.
\]
We can omit the limited liability constraints in the optimization problem. Since all interim payments are set to zero, the limited liability constraints are satisfied by construction at those stages. Moreover, the final payment must be nonnegative because it is guaranteed by the IR constraint in the last period.

\begin{lemma}[Payoff Equivalence]
\label{lem:utexpression}
For any incentive-compatible sequence of action rules
$\{x_t\}_{t=1}^{N}$, define
\begin{align*}
\hat{u}_t(h_t) \equiv{}& \int_{\theta_1}^{\bar{\theta}} x_1(s)\,ds +\sum_{i=1}^{t-1}\theta_i x_i(h_i) +\sum_{i=2}^{t} \left[ \int_{\theta_i}^{\bar{\theta}} x_i(h_{i-1},s)\,ds - \int_{\underline{\theta}}^{\bar{\theta}} F(s)x_i(h_{i-1},s)\,ds \right].
\end{align*}
Then, for every $t\in\mathcal N$ and $h_t\in\Theta^t$, $u_t(h_t) = u_1(\bar{\theta}) + \hat{u}_t(h_t)$.
\end{lemma}

Lemma \ref{lem:utexpression} shows that the agent's expected utility can be decomposed into two components: $\hat{u}_t(h_t)$, which is a function of the action rules up to period $t$, and $u_1(\bar{\theta})$, the expected utility of the highest type in the first period. This comes from repeatedly applying the envelope formula in Lemma \ref{lem:ICconditions} and connecting the expected utility across different periods. To illustrate this, consider the case when $t=2$ and the realized cost in the first period is $\theta_1$. Then we have,
\[
    u_2(\theta_1,\theta_2)=u_2(\theta_1,\bar{\theta})+\int_{\theta_2}^{\bar{\theta}}x_2(\theta_1,s)ds.
\] 
Meanwhile, from the agent's perspective in period 1, after the realization of $\theta_1$ and before the realization of $\theta_2$, his expected utility $u_1(\theta_1)$ equals the expectation of the expected payoff in period $2$ minus the cost of working in period $1$, i.e.
\begin{align*}
u_1(\theta_1)
&=
E_{\tilde{\theta}_2}
\left[
u_2(\theta_1,\tilde{\theta}_2)
\right]
-
\theta_1 x_1(\theta_1)\\
&=
\int_{\underline{\theta}}^{\bar{\theta}}
\left[
u_2(\theta_1,\bar{\theta})
+
\int_{\theta_2}^{\bar{\theta}}
x_2(\theta_1,s)\,ds
\right]
dF(\theta_2)
-
\theta_1 x_1(\theta_1)\\
&=
u_2(\theta_1,\bar{\theta})
+
\int_{\underline{\theta}}^{\bar{\theta}}
F(s)x_2(\theta_1,s)\,ds
-
\theta_1 x_1(\theta_1).
\end{align*}
Combining this with the fact that $ u_1(\theta_1)= u_1(\bar{\theta})+\int_{\theta_1}^{\bar{\theta}}x_1(s)ds$, we can express $u_2(\theta_1,\bar{\theta})$ as a function of the action rules and $u_1(\bar{\theta})$. Similarly, $u_3(h_2,\bar{\theta})$ is a function of the action rules and $u_2(h_1,\bar{\theta})$, which itself is a function of the action rules and $u_1(\bar{\theta})$. This pattern continues, allowing us to express $u_t(h_{t-1},\bar{\theta})$ as a function of the action rules and $u_1(\bar{\theta})$ for all $t$ and $h_{t-1}$. Note that expressing $u_t$ as a function of the action rules and $u_1(\bar{\theta})$ is standard in dynamic mechanism design. What is nonstandard in our setting is that $u_1(\bar{\theta})$ is not generally equal to zero.\footnote{In most of the existing literature on static mechanism design and dynamic mechanism design without limited liability (see \cite{pavan2014dynamic}), $u_1(\bar{\theta})$ can be normalized to zero without loss, as this remains implementable and does not grant the agent additional surplus.} Consequently, determining $u_1(\bar{\theta})$ becomes a nontrivial step in the analysis. To illustrate, consider the example in Figure \ref{fig:u1example}.

\begin{figure}[h]
            \centering
            \begin{tikzpicture}[x= 5.5cm, y = 5.5cm]
                \node at (-0.3, 0.1) {Types};
                \node at (0, 0.6) {$t = 1$};
                \node at (0.3, 0.6) {$t = 2$};

                \node at (0, 0.4) {0.5};
                \node at (0, 0.2) {0.5};
                \node at (0, 0) {1.5};
                \node at (0, -0.2) {1.5};
                
                \node at (0.3, 0.4) {0.5};
                \node at (0.3, 0.2) {1.5};
                \node at (0.3, 0) {0.5};
                \node at (0.3, -0.2) {1.5};

                \draw[blue, thick] (0, 0.4) circle (10pt);
                \draw[blue, thick] (0, 0.2) circle (10pt);
                \draw[blue, thick] (0.3, 0.4) circle (10pt);
                \draw[blue, thick] (0.3, 0.2) circle (10pt);
                \draw[blue, thick] (0.3, 0) circle (10pt);
                \draw[blue, thick] (0.3, -0.2) circle (10pt);

                \node[red,right] at (0.45,0.4) {\$2};
                \node[red, right] at (0.45,0.2) {\$2};
                \node[red, right] at (0.45,0) {\$1.5};
                \node[red, right] at (0.45,-0.2) {\$1.5};
                \end{tikzpicture}
                \caption{An example for explaining $u_1(\bar{\theta}) \neq 0$.
                }
                \label{fig:u1example}
        \end{figure}

 The agent has costs $0.5$ and $1.5$ with equal probability.\footnote{The two-point distribution falls outside the model's
continuous-density assumption; we use it only for simplicity, and the
same logic applies under a continuous density.} Consider the action rule that requires the low-cost type to work in the first period and both types to work in the second period. The corresponding optimal payment rule is to pay the agent $\$2$ if he works in both periods and $\$1.5$ if he works only in the second period. In this example, 
$u_1(\bar{\theta}) = \tfrac{1}{2}(1.5 - 1.5) + \tfrac{1}{2}(1.5 - 0.5) = 0.5$.
If instead we were to impose $u_1(\bar{\theta}) = 0$, then this action rule would no longer be implementable. This restriction is therefore not without loss, as it shrinks the feasible set of action rules and leads to a suboptimal mechanism. Indeed, in the optimal mechanism, $u_1(\bar{\theta})$ is strictly positive.

Lemma \ref{lem:u1expression} characterizes the optimal choice of $u_1(\bar{\theta})$ as the smallest value that ensures the mechanism is implementable.

\begin{lemma} \label{lem:u1expression}
Fix any incentive-compatible action rules
$\{x_t\}_{t=1}^{N}$. The periodic IR constraints are equivalent to: $u_1(\bar{\theta}) + \hat{u}_t(h_{t-1},\bar{\theta}) \geq 0, \; \forall\, t\in\mathcal N,\; h_{t-1}\in\Theta^{t-1}$. Among all values satisfying these constraints, the principal optimally chooses the smallest one. Hence,
\begin{align*}
u_1^*(\bar{\theta}) &= \inf\Bigl\{ u\in\mathbb R: u+\hat{u}_t(h_{t-1},\bar{\theta})\geq0,\; \forall\,t\in\mathcal N,\; h_{t-1}\in\Theta^{t-1} \Bigr\} \\
&= -\inf_{\substack{ t\in\mathcal N \\ h_{t-1}\in\Theta^{t-1} }} \hat{u}_t(h_{t-1},\bar{\theta}).
\end{align*}
\end{lemma}

Here, the principal must choose $u_1(\bar{\theta})$ to guarantee the periodic IR constraints hold everywhere. Lemma \ref{lem:utexpression} shows that $u_t = u_1(\bar{\theta}) + \hat{u}_t$. Therefore, the smallest $u_1(\bar{\theta})$ that ensures the mechanism is implementable is the infimum of $u_1(\bar{\theta})$ such that $u_1(\bar{\theta})+\hat{u}_t(h_t) \geq 0$ for all $t$ and $h_t$. Any value smaller than this would violate some periodic IR constraint, rendering the mechanism not implementable.

\begin{proposition}[Revenue Equivalence]
\label{prop:revenueequivalence}
For any incentive-compatible sequence of action rules
$\{x_t\}_{t=1}^{N}$, the terminal payment rule that implements these
action rules and maximizes the principal's expected payoff is uniquely
determined by
\begin{align*}
p_N(h_N)
={}& u_1^*(\bar{\theta})
+\int_{\theta_1}^{\bar{\theta}} x_1(s)\,ds
+\sum_{i=1}^N \theta_i x_i(h_i) \\
&+\sum_{i=2}^N
\left[
\int_{\theta_i}^{\bar{\theta}} x_i(h_{i-1},s)\,ds
-\int_{\underline{\theta}}^{\bar{\theta}}
F(s)x_i(h_{i-1},s)\,ds
\right],
\end{align*}

where $u_1^*(\bar{\theta})$ is given by
Lemma~\ref{lem:u1expression}.
\end{proposition}

Combining the fact that the agent's expected utility in the last period can be expressed as $u_N(h_N)=\hat{u}_N(h_N) + u_1^{*}(\bar{\theta})$ and the fact that $u_N(h_N)= p_N(h_N) - \theta_N \cdot x_N(h_N)$, we can express $p_N$ as a function of the action rules and $u_1^{*}(\bar{\theta})$, which itself is determined by the action rules. Proposition \ref{prop:revenueequivalence} provides such a formula for $p_N$. As a result, the principal's optimization problem can be simplified as an optimization problem over cutoffs.

\subsection{Reduce dimensionality} \label{sec:reducedim}

To make the principal's optimization problem more tractable, Proposition \ref{prop:summarystats} shows that the agent's working status history can serve as a sufficient statistic for the cost history in determining the future cutoffs. Accordingly, throughout the analysis, we will focus on mechanisms whose cutoffs depend only on the working status history. This reduction in dimensionality shifts the optimization problem from addressing uncountably many cutoffs to optimizing over finitely many cutoffs.

\begin{proposition}
\label{prop:summarystats}
Without loss of optimality, the period-$t$ cutoff can be chosen to
depend on the cost history $h_{t-1}$ only through the working-status
history $w_{t-1}$. Thus, the principal may restrict attention to
cutoff rules of the form $c_t(h_{t-1})=c_t(w_{t-1})$ for every $t\in\mathcal N$.
\end{proposition}

The intuition behind this proposition is as follows. First, the principal's payoff from each period is additively separable. Second, since costs are i.i.d., historical costs do not provide any information about future costs.  Because of these two features, one might think that the future cutoff selection is independent of past costs. However, there is one (and only one) link between them: the accumulated promised yet postponed payment that the principal owes the agent for working in the past.\footnote{This is the part of the final payment that depends on past cost realizations and past working status.} It affects the future cutoff selection as it provides the principal with negotiation power to discipline the agent. Since the accumulated due payments depend only on the working status history given the cutoff rules, it suffices to condition the future cutoffs on the working status history only.\footnote{If the action rules were not cutoff rules, such as if the action space were continuous, $w_t$ would still be a sufficient summary statistic of $\theta_t$, although it might not simplify the problem as much for continuous $x_t$.}

Technically speaking, 
past cost history determines the accumulated promised payment, which in turn determines the range of future cutoffs that satisfy the IR constraints. The future cutoffs after the same working status history appear symmetrically in the objective function, implying that their optimal values should be the same.

Proposition \ref{prop:summarystats} reduces the number of cutoffs to a finite number. In each period \( t \), the number of cutoffs equals the number of possible working histories, i.e., \( 2^{t-1} \). Hence, the total number of cutoffs is
$
2^0 + 2^1 + \ldots + 2^{N-1} = 2^N - 1.
$
For example, with 3 periods, the problem reduces to choosing 7 cutoffs:
$
c_1, c_2(1), c_2(0), c_3(1,1), c_3(1,0), c_3(0,1), c_3(0,0)
$
in \( \Theta^7 \). By Proposition \ref{prop:revenueequivalence}, under cutoff rules, the payment rule $p_N(w_N)$ becomes:
\begin{align}\label{eq:paymentrule1}
    p_N(w_N) = \sum_{t=1}^{N} x_t\cdot c_t(w_{t-1})- \sum_{t=2}^{N} \int_{\underline{\theta}}^{c_t(w_{t-1})} F(\theta)d\theta + u^{*}_1(\bar{\theta}).
\end{align}

Substituting \eqref{eq:paymentrule1} into the principal's payoff, we can rewrite the optimization problem as:
\begin{align} \label{equ:optimprob1}
     \max_{\{c_t(w_{t-1})\}_{t=1}^N} E_{\pmb{\theta}} \left[\alpha \sum_{t=1}^{N} x_t - \sum_{t=1}^{N} x_t\cdot c_t(w_{t-1}) +\sum_{t=2}^{N} \int_{\underline{\theta}}^{c_t(w_{t-1})} F(\theta)d\theta - u^{*}_1(\bar{\theta}) \right].
\end{align}

Note that Proposition \ref{prop:summarystats} also reduces the number of IR constraints to a finite number, transforming $u^{*}_1(\bar{\theta})$ from an infimum function to a minimum function.\footnote{\label{fn:auxiliary}
Some binary working-status histories are not induced by any report
history. We call these \emph{auxiliary histories} and set all cutoffs
at auxiliary histories to $\underline{\theta}$. This is without
loss: an auxiliary history begins with an impossible shirking status,
and the normalization sets all subsequent integral terms to zero.
Hence, no auxiliary history affects the minimum defining
$u_1^*(\bar{\theta})$. Since auxiliary histories cannot be induced by any report history,
this normalization affects neither the induced allocation nor the
principal's expected payoff.}
\begin{align*}
    u^{*}_1(\bar{\theta})
    &= -\inf_{h_{t-1} \in \Theta^{t-1}, t\in \mathcal{N}}  \hat{u}_t(h_{t-1},\bar{\theta})\\
    &= - \min_{w_{t-1} \in \{0,1\}^{t-1}, t\in \mathcal{N}}  \left( \sum_{i=1}^{t-1} x_i\cdot c_i(w_{i-1}) - \sum_{i=2}^{t} \int_{\underline{\theta}}^{c_i(w_{i-1})} F(\theta)d\theta   \right).
\end{align*}

We can now omit the periodic IC and IR constraints from the optimization problem. The IC constraints are satisfied by the cutoff rules and the formula for the final payment, while the IR constraints are ensured by the optimal choice of \( u_1^*(\bar{\theta}) \). The problem therefore reduces to an unconstrained optimization over the cutoffs. Despite this simplification, the problem remains nontrivial. The objective function includes \( u_1^*(\bar{\theta}) \), which is defined as a minimum function, giving rise to a nested optimization problem: an inner minimization and an outer maximization. We address this problem in the next section.

\subsection{Solve the nested optimization problem} \label{sec:simplifyoptim}

To tackle the nested optimization problem in \eqref{equ:optimprob1}, we will first solve the inner minimization problem. Specifically, we find the IR constraint that must bind and determine the value of $u^{*}_1(\bar{\theta})$ accordingly. We then add constraints to ensure that all other IR constraints are satisfied. This allows us to reformulate the nested optimization problem as a standard optimization problem with constraints.

\begin{lemma} \label{lem:optmechproperties}
 An optimal mechanism exists. Moreover, every optimal mechanism is either the always-working mechanism or has interior initiation cutoffs: $c_t(\mathbf{0}) \in (\underline{\theta},\bar{\theta})$ for all $t \in \mathcal{N}$. 
\end{lemma}

 Lemma \ref{lem:optmechproperties} has two ingredients. First, all cutoffs must be strictly greater than $\underline{\theta}$. Intuitively, this follows from the efficiency of working.  If the agent always shirks in some period, then we show that the principal can be strictly better off by inducing the agent to work with a small positive probability in that period, regardless of whether this change affects \(u_1^*(\bar{\theta})\).

Second, if the principal intends to induce the agent to work for all cost realizations in some period \(t\), conditional on no prior work, then she should do so in the first period. The reason is that the principal's expected payoff from setting \(c_t(\mathbf{0})=\bar{\theta}\) decreases with \(t\). Because the principal can achieve the first-best outcome in all subsequent periods, setting \(c_t(\mathbf{0})=\bar{\theta}\) in an earlier period allows her to implement the first best in more periods. Therefore, we have either \(c_1(\mathbf{0})=\bar{\theta}\), in which case the optimal mechanism induces the agent to work in every period, or all initiation cutoffs being interior.

The first case is straightforward and corresponds to the always-working mechanism. For the second case, we need to identify the IR constraint that must bind to determine the value of \( u_1^*(\bar{\theta}) \). Specifically, we show that the IR constraint for the history in which the agent shirks in all \( N \) periods, namely \( w_N = \mathbf{0}_N \)\footnote{Lemma \ref{lem:optmechproperties} guarantees that \( \mathbf{0}_N \) is an on-path history.}, must bind. Otherwise, the principal could strictly increase her payoff by slightly raising the cutoff \( c_N(\mathbf{0}_{N-1}) \) while preserving all other IR constraints.\footnote{The proof of this statement is provided in the proof of Proposition \ref{prop:constrainedoptim}.} Thus,
\[
u_1^*(\bar{\theta}) + \hat{u}_N(\mathbf{0}_N) = 0
\qquad
\text{(IR constraint for } w_N = \mathbf{0}_N \text{)}.
\]
This condition allows us to solve the inner minimization problem by replacing \( u_1^*(\bar{\theta}) \) with
$
-\hat{u}_N(\mathbf{0}_N)
=
\sum_{t=2}^{N}
\int_{\underline{\theta}}^{c_t(\mathbf{0})}
F(\theta)\, d\theta.
$
At the same time, we must impose constraints ensuring that all other IR constraints are satisfied. These constraints also guarantee that \( \hat{u}_N(\mathbf{0}_N) \) is indeed the minimum among $\hat{u}_t(w_t)$ for all $t$ and $w_t$. Therefore, the nested optimization problem can be reformulated as the following standard constrained optimization problem:

\begin{proposition} \label{prop:constrainedoptim}
   If an optimal mechanism has interior initiation cutoffs, then it must be a solution to the following constrained optimization problem\footnote{By Lemma \ref{lem:optmechproperties}, if the optimization problem \eqref{equ:optimprob2} does not admit a solution, then the optimal mechanism is the always-working mechanism, given that an optimal mechanism exists.}:

  {
    \small
          \begin{align} \label{equ:optimprob2}
        &  \max_{\substack{
            \{c_t(w_{t-1})\}_{t=1}^{N}\\[0.8mm]
            c_t(\mathbf{0})\in
            (\underline{\theta},\bar{\theta}) \:\forall t 
        }} E_{\pmb{\theta}}  \Bigg[\alpha  \sum_{t=1}^{N} x_t - \sum_{t=1}^{N}  x_t c_t(w_{t-1})+ \sum_{t=2}^{N} \int_{\underline{\theta}}^{c_t(w_{t-1})} F(\theta)d\theta - \sum_{t=2}^{N} \int_{\underline{\theta}}^{c_t(\mathbf{0})}F(\theta)d\theta \Bigg] \nonumber\\[1mm]
        &\text{s.t.} \sum_{i=1}^{t-1} x_i c_i(w_{i-1}) - \sum_{i=2}^{t} \int_{\underline{\theta}}^{c_i(w_{i-1})} F(\theta)d\theta  \geq - \sum_{i=2}^{N} \int_{\underline{\theta}}^{c_i(\mathbf{0})}F(\theta)d\theta, \;  t \in \mathcal{N},  w_{t-1} \in \{0,1\}^{t-1}.
    \end{align}
  }
\end{proposition}

To solve the constrained optimization problem \eqref{equ:optimprob2}, we first solve the relaxed optimization problem by ignoring all the IR constraints in \eqref{equ:optimprob2}. We then find conditions under which the relaxed solution is feasible and therefore is also the solution to the original problem. 

If the relaxed problem admits a solution, it must be a consecutive-working menu: the agent starts working in each period with strictly positive probability, and once he starts, he is induced to work in every remaining period, as illustrated in Figure \ref{fig:IRconstraints}. Formally, $c_t(w_{t-1})=\bar{\theta}$ for all \(w_{t-1}\neq \mathbf{0}_{t-1}\) and \(w_{t-1}\neq\emptyset\). The reason is that these cutoffs do not enter \(u_1^*(\bar{\theta})\) and therefore do not affect the information rent or the principal's payoff from other histories. Since work is efficient as \(\alpha>\bar{\theta}\), it is optimal to set these cutoffs equal to \(\bar{\theta}\).

Next, we characterize conditions under which the relaxed solution is feasible, i.e., all IR constraints are satisfied.  We partition the work histories into two categories. The first consists of the consistent-shirking histories, \(w_t=\mathbf{0}_t\) for all \(t\in\mathcal{N}\). The second consists of all remaining histories. Figure \ref{fig:IRconstraints} displays all possible work histories using a tree structure, where each branch indicates the agent's prescribed actions in that period given the work history and current cost realization. For each branch, it suffices to verify the IR constraint at its highest node because, holding the work history fixed, the agent's expected payoff is minimized when the current cost is \(\bar{\theta}\).\footnote{Accordingly, by the IR constraint associated with a branch, we mean the constraint at its highest node, corresponding to the current cost realization \(\bar{\theta}\).}

\begin{figure}[h]
	\centering
	\begin{tikzpicture}[x= 7cm, y = 7cm][h]
        
        \node at (-0.03, 1) {\small Period 1};
        
        \draw[thick] (0, -0.08) -- (0, 0.2);
        \draw (-0.02, 0.2) -- (0.02, 0.2);
        \draw (-0.02, -0.08) -- (0.02, -0.08);
        \node[above] at (0, 0.21) {\small$\overline{\theta}$}; 
        \node[below] at (0, -0.09) {\small $\underline{\theta}$}; 
        \node[left, red] at (0, 0.15) {\small shirk}; 
        \node[left, blue] at (0, -0.0) {\small work}; 
         \filldraw[red, opacity = 0.5] (0, 0.2) circle (6pt);

        \draw[thick] (-0.01, 0.11) -- (0.01, 0.11);
        \node[right] at (0,0.1) {\small$c_{1}$};

        \draw [-Stealth, red] (0.05, 0.15) to (0.2, 0.25);
        \draw [-Stealth, blue] (0.05, 0) to (0.2, -0.1);

        \node at (0.3, 1) {\small 2};

        \draw[thick] (0.3, 0.11) -- (0.3, 0.4); 
        \draw (0.28, 0.4) -- (0.32, 0.4);
        \draw (0.28, 0.11) -- (0.32, 0.11);
        \node[left, red] at (0.3, 0.32) {\small s}; 
        \node[left, blue] at (0.3, 0.18) {\small w};
        \draw[thick] (0.29, 0.27) -- (0.31, 0.27);
        \node[right] at (0.3, 0.3) {\small $c_{2}(0)$};
        \filldraw[red, opacity = 0.5] (0.3, 0.4) circle (6pt);

        \draw[thick] (0.3, 0) -- (0.3, -0.25); 
        \draw (0.28, 0) -- (0.32, 0);
        \draw (0.28, -0.25) -- (0.32, -0.25);
        \filldraw[black] (0.3, 0) circle (1.5pt) node[anchor= west]{\small $c_{2}(1)$};
        \node[left, blue] at (0.3, -0.14) {\small w}; 
        
        \node at (0.6, 1) {\small$\cdots$};
        \node at (0.6, 0.45) {\reflectbox{$\ddots$}};
        \node at (0.6, 0.1) {$\cdots$};
        \node at (0.6, -0.15) {$\cdots$};        

        \node at (0.9, 1) {\small $N-1$};

        \draw[thick] (0.9, 0.44) -- (0.9, 0.72); 
        \draw (0.88, 0.44) -- (0.92, 0.44);
        \draw (0.88, 0.72) -- (0.92, 0.72);
        \node[left, red] at (0.9, 0.62) {\small s}; 
        \node[left, blue] at (0.9, 0.48) {\small w}; 
        \filldraw[red, opacity = 0.5] (0.9, 0.72) circle (6pt);

        \draw[thick] (0.89, 0.54) -- (0.91, 0.54);
        \node[right] at (0.9, 0.54) {\small $c_{N-1}(\mathbf{0})$};
        
        \node at (1.2, 1) {\small $N$};

        \draw [-Stealth, red] (1, 0.63) to (1.1, 0.7);
        
        \draw[thick] (1.2, 0.62) -- (1.2, 0.88); 
        \draw (1.18, 0.62) -- (1.22, 0.62);
        \draw (1.18, 0.88) -- (1.22, 0.88);
        \node[left, red] at (1.2, 0.78) {\small s}; 
        \node[left, blue] at (1.2, 0.66) {\small w}; 
        \filldraw[red, opacity = 0.5] (1.2, 0.88) circle (6pt);

        \draw[thick] (1.19, 0.7) -- (1.21, 0.7);
        \node[right] at (1.2, 0.72) {\small $c_{N}(\mathbf{0})$};
    
        \draw [-Stealth, blue] (1, 0.47) to (1.1, 0.41);
        
        \draw[thick] (1.2, 0.28) -- (1.2, 0.52); 
        \draw (1.18, 0.28) -- (1.22, 0.28);
        \draw (1.18, 0.52) -- (1.22, 0.52);
        \node [left, blue] at (1.2, 0.39) {\small w}; 
        \filldraw[black] (1.2, 0.52) circle (1.5pt) node[anchor= west]{\small $c_{N}(\mathbf{0}_{N-2},1)$};

        \node at (1.2, 0.16) {$\vdots$}; 
        
        \draw [-Stealth, blue] (1, -0.15) to (1.1, -0.15);
        
        \draw[thick] (1.2, 0) -- (1.2, -0.25); 
        \draw (1.18, 0) -- (1.22, 0);
        \draw (1.18, -0.25) -- (1.22, -0.25);
        \node[left, blue] at (1.2, -0.14) {\small w}; 

        \draw[thick] (1.18, 0) -- (1.21, 0);
        \node[right] at (1.21, 0) {\small $c_{N}(\mathbf{1})$};

        \draw [blue, thick, opacity = 0.3, line width = 1 mm ] (1.3, 0.6) -- (1.4, 0.4) -- (0.2, -0.3) -- (0.1, -0.1) -- cycle;          
        \end{tikzpicture} 
        \caption{Work histories and relevant IR constraints.}
        \label{fig:IRconstraints}
\end{figure}

The consistent-shirking histories in the first category are marked by large dots. Among these histories, \( w_N=\mathbf{0}_N \) yields the lowest expected payoff for the agent.\footnote{One can verify that \(u_t(\mathbf{0}_t)\) is decreasing in \(t\). Intuitively, as time passes, the agent has fewer periods to start working and earn information rents.} The IR constraint for this work history holds by construction, as \(u_1^*(\bar{\theta})\) is chosen so that it binds. It follows that the IR constraints for all other histories in this category are also satisfied.

The second category consists of all remaining branches. Among them, the IR constraints for branches in the boxed region are relatively harder to satisfy, as they correspond to the first period in which the agent is prescribed to work for all cost realizations. If the agent is willing to participate even when his cost is \( \bar{\theta} \) at these branches, then he will also be willing to work for all cost realizations in any future periods. This is because the final payment is fixed at the end, while the costs incurred in earlier periods are sunk. Equivalently, the IR constraint for the history \( (\mathbf{0}_t,\mathbf{1}_2) \) implies the IR constraints for all histories \( (\mathbf{0}_t,\mathbf{1}_m) \) with \( m>2 \).\footnote{The IR constraint for the history \( (\mathbf{0}_t,\mathbf{1}_1) \) is implied by the IR constraint for the history \( \mathbf{0}_{t+1} \). Intuitively, the agent must receive sufficient information rents to begin working in period \( t+1 \) rather than continue shirking. Hence, these IR constraints are automatically satisfied once the IR constraints for the first category hold.}

The IR constraints for work histories in the blue box, i.e., $(\mathbf{0}_t,\mathbf{1}_2)$, are $c_{t+1}(\mathbf{0}) + \sum_{i=t+2}^{N}\int_{\underline{\theta}}^{c_i(\mathbf{0})}F(\theta)d\theta + E(\theta) - \bar{\theta} \geq 0, \text{ for } 0 \leq t \leq N-2$.
Equivalently,
\begin{equation} \label{equ:feasbilitycondition1}
\min_{0 \leq t \leq N-2} \left(c_{t+1}(\mathbf{0}) + \sum_{i=t+2}^{N}\int_{\underline{\theta}}^{c_i(\mathbf{0})}F(\theta)d\theta\right) + E(\theta) - \bar{\theta} \geq 0.
\end{equation}
Since all the initiation cutoffs are interior and the integral terms are positive, the first term in \eqref{equ:feasbilitycondition1} is strictly greater than $\underline{\theta}$:
\begin{align*} 
    \min_{0 \leq t \leq N-2} \left(c_{t+1}(\mathbf{0}) + \sum_{i=t+2}^{N}\int_{\underline{\theta}}^{c_i(\mathbf{0})}F(\theta)d\theta\right) >  \min_{t} c_{t+1}(\mathbf{0}) > \underline{\theta}.
\end{align*}
Therefore, we derive a sufficient condition for \eqref{equ:feasbilitycondition1}:
\[
\underline{\theta} + E(\theta) \geq \bar{\theta} \tag{Assumption \ref{assumption1}}.
\]
It follows that, under Assumption \ref{assumption1}, the solution to the relaxed optimization problem is feasible, and therefore it is also the solution to the original constrained optimization problem.

In fact, Assumption~\ref{assumption1} guarantees the feasibility of every consecutive-working menu, not only that of the relaxed optimum, because the IR constraints in \eqref{equ:feasbilitycondition1} are satisfied for any vector of interior initiation cutoffs.  The conclusion of Theorem~\ref{thm:optimalmechproperties} therefore continues to hold under the weaker sufficient condition that guarantees only the feasibility of the relaxed optimum. Specifically, suppose the relaxed problem of \eqref{equ:optimprob2} admits a solution, and let $\{c^*_t(\mathbf{0})\}_{t=1}^N$ denote its initiation cutoffs.\footnote{After setting all post-initiation cutoffs equal to \(\bar{\theta}\), the relaxed
problem reduces to a smooth \(N\)-dimensional optimization over the
initiation cutoffs \(\{c_t(\mathbf{0})\}_{t=1}^N\). Since any solution is interior, first-order conditions can be used to identify candidate solutions.} It is then sufficient for \eqref{equ:feasbilitycondition1} to hold at ${c_t^*(\mathbf{0})}_{t=1}^N$ to conclude that an optimal mechanism is either a consecutive-working menu or the always-working mechanism.\footnote{Under Assumption~\ref{assumption1}, the relaxed problem of \eqref{equ:optimprob2} is equivalent to the original constrained problem. Therefore, if the relaxed problem does not admit a solution, the optimal mechanism must be the always-working mechanism.}

To understand the intuition behind the optimal mechanism from a technical perspective, consider the payment rule while temporarily ignoring the term \(u_1^*(\bar{\theta})\). The remaining component that depends on \(c_t(w_{t-1})\) can be interpreted as the payment promised for working in period \(t\) following history \(w_{t-1}\). Under the cutoff \(c_t\), the agent's expected cost of working is
$
F(c_t)c_t-\int_{\underline{\theta}}^{c_t}F(\theta)\,d\theta,
$
which is exactly the part of the payment that depends on $c_t$ besides \(u_1^*(\bar{\theta})\). Thus, if the principal can set \(c_t=\bar{\theta}\) without increasing \(u_1^*(\bar{\theta})\), she induces the agent to work for all cost realizations while paying only the expected cost, thereby achieving the first-best outcome in that period. Assumption \ref{assumption1} guarantees precisely that assigning the agent to work for all cost realizations in periods after he starts working will not affect \(u_1^*(\bar{\theta})\). Consequently, the principal achieves the first-best outcome in all subsequent periods once the agent begins working.

\section{Properties of the Optimal Mechanism} \label{sec:properties}

In this section, we first discuss the condition that determines whether the optimal mechanism is a consecutive-working menu or an always-working mechanism in Section \ref{sec:comparativestatics}. We then characterize the structure of the general optimal mechanism without Assumption \ref{assumption1} in Section \ref{sec:generaloptimalmech}. Finally, in Section \ref{sec:backloading}, we discuss the role of backloading payments, emphasizing that the optimal mechanism requires only part of the payment to be backloaded, while the remainder can be paid as early as the second period in which the agent works with certainty.

\subsection{Comparative statics of the optimal mechanism} \label{sec:comparativestatics}

We now study how the principal's value of work determines whether
the optimal mechanism is a consecutive-working menu or the
always-working mechanism.

\begin{theorem} \label{thm:mainresult}
Under Assumption \ref{assumption1}, there exists
\(\hat{\alpha}\geq\bar{\theta}\) such that a consecutive-working menu
is optimal for
\(\bar{\theta}\leq\alpha<\hat{\alpha}\), whereas the always-working
mechanism is optimal for \(\alpha\geq\hat{\alpha}\).
\end{theorem}

Let us illustrate Theorem \ref{thm:mainresult} using Figure \ref{fig:comparison}. Consider the mechanism that maximizes the principal's expected payoff within the class of not-always-working mechanisms, i.e., mechanisms that allow the agent to shirk in at least one period.\footnote{ Because this class is not compact, an optimum in this set may not exist. Under Assumption \ref{assumption1}, it exists for $\alpha<\hat{\alpha}$ but does not exist for $\alpha>\hat{\alpha}$, as illustrated in Figure \ref{fig:comparison}.} Whenever it exists, we refer to it as the optimal not-always-working mechanism. Since an optimal mechanism exists by Lemma \ref{lem:optmechproperties}, it must be either the optimal not-always-working mechanism or the always-working mechanism. 

In Figure~\ref{fig:comparison}, the curve represents the principal's expected payoff under the optimal not-always-working mechanism and the straight line under the always-working mechanism. The curve lies above the straight line at $\alpha = \bar{\theta}$. By continuity, there exists a region to the right, $\bar{\theta} \leq \alpha < \hat{\alpha}$, where the optimal not-always-working mechanism is the optimal mechanism. Under Assumption~\ref{assumption1}, it takes the form of a consecutive-working menu in this region, as highlighted by the thick curve. For $\alpha \geq \hat{\alpha}$, the optimal mechanism switches to the always-working mechanism. We show that the always-working mechanism becomes relatively more attractive as $\alpha$ increases, so there is only one crossing point: once the optimal mechanism switches to always-working, it never switches back.

\begin{figure}[h]
    \centering
        \begin{tikzpicture}[x = 6cm, y = 6cm]
            
            \draw [->] (0,0) -- (1.1,0) node[right] {$ \alpha $}; 
            \draw [->] (0,-0.4) -- (0,0.9) node[above] {$ V $};

                \draw [black,thick] (0, -0.3) -- (1, 0.7);
                \node [right, black] at (1, 0.7) {always-working mechanism};

                \draw[domain=0:0.8, smooth, thick, variable=\x, blue] plot ({\x},{.654*\x^(1.2)});  
                \node [right, red] at (0.85, 0.5) {consecutive-working menu};

                \draw [dashed] (0.8, 0) -- (0.8, 0.5);
                \filldraw [red] (0.8, 0.5) circle (1pt);
                \node [below] at (0.8,0) {$\hat{\alpha}$};   
                \node [below] at (0.025,0) {$\underline{\theta}$};   

                \draw[domain=0.5:0.8, smooth, variable=\x, red, line width = 1 mm, opacity = 0.5] plot ({\x},{.654*\x^(1.2)});  
                \node [above, blue] at (0.35, 0.4) {not-always-working};
                
                \draw [dashed] (0.5,0) -- (0.5, 0.285);
                \node [below] at (0.5, 0) {$\overline{\theta}$};
        \end{tikzpicture}
        \caption{Principal's expected payoffs under different mechanisms.}
        \label{fig:comparison}
\end{figure}

\subsection{General optimal mechanism} \label{sec:generaloptimalmech}

We now characterize the structure of the general optimal mechanism without Assumption~\ref{assumption1}. Theorem~\ref{thm:generaloptimalmech} shows that consecutive working is a robust feature of the optimal mechanism.

\begin{theorem} \label{thm:generaloptimalmech}
   The optimal mechanism has a two-stage structure: for every
on-path work history $w_N$, there exists
$t(w_N)\in \{0,1,\ldots,N\}$ such that
    \begin{enumerate}[(i)]
        \item Stage 1: $c_i(w_{N}^{1:i-1}) \in (\underline{\theta}, \bar{\theta})$ for all $i \leq t(w_N)$ (work at low costs and shirk at high costs, accumulating deposit).
        \item Stage 2: $c_i(w_{N}^{1:i-1}) = \bar{\theta}$ for all $i > t(w_N)$ (work for sure in all subsequent periods).
    \end{enumerate}
Here, $w_{N}^{1:i-1}$ denotes the first $i-1$ elements of the work history $w_N$, with $w_N^{1:0}=\emptyset$.
\end{theorem}

The general optimal mechanism takes the form of a two-stage mechanism. In the first stage, the agent is prescribed to work when costs are low and allowed to shirk when costs are high. The purpose of the first stage is to accumulate the ``deposit'' (promised but backloaded payment). Once the accumulated ``deposit'' becomes large enough, the mechanism enters the second stage,  in which the agent is prescribed to work consecutively until the end. In fact, as discussed in Section \ref{sec:backloading}, the principal can begin paying the agent in the second stage, as long as the ``deposit'' accumulated in the first stage remains deferred until the end. Thus, one can interpret the first stage as a trial period, during which the agent is asked to work only when costs are low. He is then officially hired in the second stage, where he is required to work in every remaining period, and the principal can also begin paying him regularly, for example, the expected cost $E(\theta)$ in each period.

The consecutive-working menu is a special case of the two-stage mechanism. Under Assumption \ref{assumption1}, the first stage collapses to a single period, so that once the agent starts working, the mechanism immediately enters the second stage. In general, the first stage can last for multiple working periods. Its length varies with the work history, as the endpoint of the first stage, $t(w_N)$ is endogenous and depends on the work history $w_N$. Thus, the two-stage mechanism distorts both the starting period and the actions in the first stage, but not the actions in the second stage.

\subsection{Discussion of payment backloading} \label{sec:backloading}

In the main analysis, we focus on the case in which all payments are backloaded to the end. This is without loss of generality, but it is not a requirement for the optimal mechanism. 
In fact, the principal can begin paying the agent in the second stage, as long as the promised payment accumulated in the first stage remains on hold until the end. For example, since the agent earns the expected cost $E(\theta)$ in every period during the second stage, the principal can pay the agent up to this amount in each period. More generally, the principal can pay the agent in an arbitrary manner for the compensation already earned in the second stage, as long as the ``deposit'' kept on hold is at least as large as the total promised payment accumulated in the first stage. For the consecutive-working menu and the always-working mechanism, payments may begin as early as the second working period, since holding the promised payment for initiating work is sufficient to ensure the agent's continued participation.

\section{Extension: Stochastic Mechanisms} \label{sec:extensions}

In this section, we consider stochastic mechanisms, which play a very different role here than in a static environment. In a static setting, randomization can often be viewed as simply convexifying the action space: for example, asking the agent to work with probability $0.5$ is equivalent to asking him to work for half of the time. In our dynamic setting, however, 
the two realizations of a randomized action can lead to distinct continuation states and therefore cannot be collapsed into a single intermediate action. Randomization instead serves as a bridge between these two states, allowing the principal to redistribute information rent across them. We show that this cross-state reallocation can lead to a stochastic improvement over the optimal deterministic mechanism.

\begin{proposition} \label{prop:stochasticimprov}
    (Stochastic Improvement) Under Assumption \ref{assumption1}, if the optimal deterministic mechanism is a consecutive-working menu, then there exist stochastic mechanisms that strictly improve the principal's expected payoff.
\end{proposition}

Let us illustrate Proposition \ref{prop:stochasticimprov} using Figure \ref{fig:stochasticimprov}. The horizontal axis represents the agent's first-period cost, while the vertical axis, $q_1(\theta_1)$, represents the first-period action rule, i.e., the probability that the agent works given the cost realization $\theta_1$. Under deterministic mechanisms, $q_1(\theta_1)$ takes binary values in $\{0,1\}$, whereas under stochastic mechanisms, it can take any value in $[0,1]$. The black line represents the action rule under the optimal deterministic mechanism, which takes the form of a consecutive-working menu. As shown in the figure, this action rule is a cutoff rule: the agent is prescribed to work with certainty when $\theta_1\leq c_1^*$ and to shirk with certainty when $\theta_1>c_1^*$.

\begin{figure}[h]
	\centering
		\begin{tikzpicture}[x = 7cm, y = 7cm]
			
			\draw [->] (0,0) -- (0.8,0) ; 
			\draw [->] (0, 0) -- (0,0.8) ;

                \node [left] at (0, 0.5) {$1$};
                \node [left] at (0, 0) {$0$};
                \draw[dashed] (0,0.5) -- (0.2, 0.5);

                \node [below] at (0.2, 0) {$\underline{\theta}$};
                \node [below] at (0.4,0) {$x^{SB}$};
                \node [below] at (0.55,0) {$c_{1}^{*}$};
                \node [below] at (0.7,0) {$\overline{\theta}$};

                \draw [dashed] (0.2, 0.5) -- (0.2,0);
                \draw [dashed] (0.4, 0.5) -- (0.4,0);
                \draw [black, thick] (0.2, 0.5) -- (0.55, 0.5);
                \draw [dashed] (0.55,0.5) -- (0.55,0);

                \draw [dashed] (0,0.45) -- (0.4,0.45);
                \draw [red, thick] (0.4,0.45) -- (0.55, 0.45);
                \node [left] at (0, 0.45) {$1-\varepsilon$};
                
                \draw[decoration={brace,mirror, raise= 2pt}, decorate] (0.55, 0.45) -- node[right= 2pt] {$\varepsilon$} (0.55, 0.5);


                \draw [->] (0,0) -- (0.8,0) node[right] {$\theta_1$}; 
                \draw [->] (0,0) -- (0,0.8) node[above] {$q_1(\theta_1)$}; 

		\end{tikzpicture}
        \caption{Stochastic improvement for action rule in the first period.}
        \label{fig:stochasticimprov}
\end{figure}

Let $x_{SB}$ denote the cutoff under the optimal static mechanism when there is only one period. As shown in the figure, $c_1^*>x_{SB}$, indicating that the principal induces the agent to work for a larger set of cost realizations in the first period when there are multiple periods. Doing so requires the principal to pay additional information rent in the first period. Why is she willing to incur this cost? The reason is that her continuation payoff is higher once the agent starts working, because she can achieve the first-best outcome in all subsequent periods. Thus, the principal is willing to pay additional information rent in the first period in order to increase the probability of reaching this more favorable continuation. We then show that a stochastic mechanism allows the principal to achieve the same continuation payoff while paying less information rent.

To see the stochastic improvement, consider modifying the action rule to the red line for $\theta_1\in[x_{SB},c_1^*]$, thereby reducing the probability of working by a small amount $\varepsilon$ for types in this interval. This change reduces the agent's information rent in the first period. More importantly, the principal can implement it without sacrificing her continuation payoff. To see this, set the agent's expected payoff conditional on working in the first period ($x_1=1$) equal to zero, and allocate all of the information rent to the realization in which the agent shirks ($x_1=0$). Specifically, when $x_1=1$, the principal promises the agent a payment of $\theta_1 + (N-1)E(\theta)$ and prescribes him to work with certainty in every subsequent period. Under Assumption \ref{assumption1}, this allows the principal to achieve the first-best outcome in all subsequent periods. When $x_1=0$, the principal instead promises the agent a payment of 
\[
    \frac{u_1^*(\bar{\theta})+\int_{\theta_1}^{\bar{\theta}}q_1(s)ds}{\varepsilon}+ (N-1)E(\theta),
\]
and again prescribes him to work with certainty in every subsequent period. The first term is the agent's first-period information rent divided by the probability of shirking, while the second term covers his expected cost of working in the remaining periods. For sufficiently small $\varepsilon$, the first term becomes large enough to induce the agent to work with certainty in every subsequent period. Hence, the principal can achieve the first-best continuation outcome following either realization of $x_1$. The stochastic mechanism therefore improves the principal's expected payoff by reducing the agent's first-period information rent while preserving the same continuation payoff. Proposition \ref{prop:stochasticimprov} formalizes this argument.

Fully characterizing the optimal stochastic mechanism is technically challenging and beyond the scope of this paper. Nevertheless, we establish one general property of the optimal stochastic mechanism: the implementation illustrated above is in fact an optimal way to implement a stochastic mechanism. Proposition~\ref{prop:stochasticimplementation} shows that an optimal continuation implementation can be chosen so that the agent receives zero continuation payoff when the action realization is work, while all of his interim utility is allocated to the shirking realization.

\begin{proposition}
\label{prop:stochasticimplementation}
Under Assumption~\ref{assumption1}, fix an on-path history
$z_{t-1}\equiv(h_{t-1},w_{t-1})$, an implementable stochastic action rule
$q_t(z_{t-1};\theta_t)$, and the associated promised
utility \(u_t(z_{t-1};\bar{\theta})\). An optimal continuation mechanism can be chosen such that, $ \forall \: \theta_t$ with
$q_t(z_{t-1};\theta_t)<1$,
\[
u_t(z_{t-1};\theta_t,x_t=1)=0 \quad \text{and} \quad  u_t(z_{t-1};\theta_t,x_t=0)=\frac{u_t(z_{t-1};\bar{\theta}) + \int_{\theta_t}^{\bar{\theta}}q_t(z_{t-1};s)\,ds}{1-q_t(z_{t-1};\theta_t)}.
\]
\end{proposition}

Here, the stochastic action rule assigns work with probability $q_t$ and shirking with probability $1-q_t$, with the realized action denoted by $x_t$. Note that both $q_t$ and the agent's expected payoff $u_t$ now depend not only on the cost history $h_t$, but also on the history of realized actions $w_{t-1}$.

There are two parts to the intuition behind the stochastic improvement. First, and more straightforwardly, randomization effectively makes the binary action space continuous, giving the principal greater flexibility by allowing some types of the agent to work only ``partially.'' Second, and more importantly, the principal generally has greater incentive power following a realization of work, because the promised future payments can serve as a deposit. By contrast, following a realization of shirking, her incentive power is weaker, since she need not promise any future payments in that state.\footnote{According to the payment rule in \eqref{eq:paymentrule1}, the accumulated promised payment following shirking in any period $t$ is in fact negative: $-\int_{\underline{\theta}}^{c_t}F(\theta) d\theta$.} A stochastic mechanism gives the principal the flexibility to reallocate promised payment across these two action realizations, thereby improving her overall continuation payoff. In particular, it allows her to transfer information rent from the state in which she has relatively abundant incentive power to the state in which incentive power is scarce.

Despite the stochastic improvement, it remains meaningful to focus on deterministic mechanisms, because stochastic mechanisms may be difficult to implement in practice. In particular, they may raise credibility concerns: It is difficult for the agent to verify whether the randomization is conducted as promised. If the agent does not fully trust the principal, as he is unable to detect deviations in the action recommendation, then the stochastic mechanism may not be implementable. Nevertheless, studying the stochastic improvement is still useful,  as it shows that the previous intuition of using past compensation to incentivize future work still applies here and can be used more flexibly and powerfully under stochastic mechanisms.

\section{Conclusion} \label{sec:conclusion}

In this paper, we study a dynamic mechanism design problem with limited liability. We show that, under certain distributional conditions, the optimal mechanism takes one of two simple forms: a consecutive-working menu or an always-working mechanism. Both can be viewed as menus of starting periods under which, once the agent starts working, he is incentivized to work consecutively until the end. The set of available starting periods is either the full set of periods or only the first period. In addition, we show that, even without any distributional assumptions, the general optimal mechanism retains a consecutive-working structure.

From traditional labor contracts to reward schemes on modern internet platforms, many mechanisms require work or consumption over consecutive periods. This paper provides an explanation from the principal's profit-maximization perspective. The consecutive-working pattern emerges in the optimal mechanism even when the agent's costs are i.i.d.\ over time and payoffs are additively separable across periods. With persistent costs, we conjecture that this pattern may become even more pronounced. Because the agent begins working when his cost is relatively low, positive serial correlation in costs makes it more likely that his cost will remain low and hence that he will continue working in subsequent periods. We leave the analysis of persistent costs for future research.

An important feature of our model is that the number of periods, $N$, is finite, which prevents us from using the stationary analyses that are commonly applied in many dynamic contracting problems. Instead, we need to characterize the optimal action rule explicitly along every history, and a key step is to determine where the participation constraint binds. In our setting, when $N$ is large, the always-working mechanism becomes approximately optimal. Intuitively, the principal can sacrifice the first period by inducing the agent to work for all cost realizations in that period, after which she can achieve the first-best outcome in every remaining period. The cost of sacrificing a single period becomes negligible as $N$ grows. Hence, the advantage of the consecutive-working menu is more salient when $N$ is small.

\begin{appendices}

\section{Proof of Results} 
\subsection{Notation for the proofs}

Let $w_{i:j}\equiv(x_i,x_{i+1},\ldots,x_j)$ denote the working-status
sequence from period $i$ to period $j$.

For any mechanism \(M\) and any history \(w_t\), let \(T_M(w_t)\) denote
the agent's expected number of future periods in which he works following
\(w_t\) under \(M\):
\[
T_M(w_t)
\equiv
\mathbb{E}^{M}\left[
    \left.
    \sum_{i=t+1}^{N}
    \mathbbm{1}\{\theta_i \leq c_i(w_{i-1})\}
    \,\right|\, w_t
\right].
\]
Let $P_M(w_t)$ denote the agent's expected future payment following $w_t$ under $M$, including the constant term $u_1^*(\bar{\theta})$. Specifically, $P_M(w_t)$ consists of the part of the final payment that does not directly depend on past cutoffs\footnote{It may still depend indirectly on past cutoffs through the constant term $u_1^*(\bar{\theta})$.}.
\[
P_M(w_t) \equiv E^M\!\left[ \left. \sum_{i=t+1}^{N} \left(x_i c_i(w_{i-1}) - \int_{\underline{\theta}}^{c_i(w_{i-1})} F(\theta)\,d\theta \right) \right|w_t \right] + u_{1,M}^*(\bar{\theta}). 
\]
We define the
principal's corresponding net continuation payoff by
\[
V_M(w_t)
\equiv
\alpha T_M(w_t)-P_M(w_t).
\]

Finally, let $V^*(\alpha)$ denote the principal's optimal expected
payoff, and let $V^{\mathrm{aw}}(\alpha)$ denote her expected payoff
under the always-working mechanism. Let $V^{cm}$ denote her expected payoff under an optimal consecutive-working menu, whenever it exists.

\subsection{Proof of Proposition \ref{prop:payatend}}

\begin{proof}
Fix any implementable mechanism $M=\{x_t,p_t\}_{t=1}^{N}$, and construct a mechanism $\bar M$ with the same action rule and $\bar p_t=0 \; \text{for } t<N, \; \bar p_N(h_N)=\sum_{i=1}^{N}p_i(h_i)$. For every $t$, $h_{t-1}$, $\theta_t$, and $\hat{\theta}_t$, the agent's continuation utility under $\bar M$ is
\[
\bar u_t(h_{t-1},\theta_t,\hat{\theta}_t)
=
u_t(h_{t-1},\theta_t,\hat{\theta}_t)
+
\sum_{i=1}^{t-1}p_i(h_i).
\]
Since the additional term is independent of the current report, all periodic IC constraints are preserved. Moreover, it is nonnegative by limited liability, so all periodic IR constraints remain satisfied. The new payment rule also satisfies limited liability, since all interim payments are zero and $\bar p_N$ is a sum of nonnegative payments. Finally, the allocation rule and the total payment at every terminal history are unchanged, so the principal's expected payoff is unchanged. Therefore, it is without loss of generality to pay the agent only in the final period.
\end{proof}

\subsection{Proof of Lemma \ref{lem:utexpression}}

\begin{proof}
We proceed by induction on $t$. For $t=1$, the envelope formula gives
\[
u_1(\theta_1)
=
u_1(\bar{\theta})
+
\int_{\theta_1}^{\bar{\theta}}x_1(s)\,ds
=
u_1(\bar{\theta})+\hat u_1(\theta_1).
\]

Now fix $t\geq2$ and suppose that $u_{t-1}(h_{t-1}) = u_1(\bar{\theta}) + \hat u_{t-1}(h_{t-1})$. Under the pay-at-the-end representation, the continuation utilities satisfy
\[
u_{t-1}(h_{t-1})
=
E_{\tilde{\theta}_t}
\left[
u_t(h_{t-1},\tilde{\theta}_t)
\right]
-
\theta_{t-1} \cdot x_{t-1}(h_{t-1}).
\]
By the envelope formula and integration by parts, we have
\[
E_{\tilde{\theta}_t}
\left[
u_t(h_{t-1},\tilde{\theta}_t)
\right]
=
u_t(h_{t-1},\bar{\theta})
+
\int_{\underline{\theta}}^{\bar{\theta}}
F(s)x_t(h_{t-1},s)\,ds.
\]
It follows that $u_t(h_{t-1},\bar{\theta})=u_{t-1}(h_{t-1})+\theta_{t-1}x_{t-1}(h_{t-1})-\int_{\underline{\theta}}^{\bar{\theta}}F(s)x_t(h_{t-1},s)\,ds$.
Applying the envelope formula once more and using the induction hypothesis yields
\begin{align*}
u_t(h_t)
={}&
u_t(h_{t-1},\bar{\theta})
+
\int_{\theta_t}^{\bar{\theta}}
x_t(h_{t-1},s)\,ds\\
={}&
u_1(\bar{\theta})
+
\hat u_{t-1}(h_{t-1})
+
\theta_{t-1}x_{t-1}(h_{t-1})+
\int_{\theta_t}^{\bar{\theta}}
x_t(h_{t-1},s)\,ds
-
\int_{\underline{\theta}}^{\bar{\theta}}
F(s)x_t(h_{t-1},s)\,ds\\
={}&
u_1(\bar{\theta})+\hat u_t(h_t),
\end{align*}
where the last equality follows from the definition of
$\hat u_t$. The result therefore holds for every $t$ and $h_t$.
\end{proof}

\subsection{Proof of Lemma \ref{lem:u1expression}}
 \begin{proof}
By Lemma~\ref{lem:utexpression}, the periodic IR constraints are equivalent to $u_1(\bar{\theta})+\hat u_t(h_t)\geq0$ for every $t\in\mathcal N$ and $h_t\in\Theta^t$.
Moreover, by the envelope formula, $u_t(h_{t-1},\theta_t)$ is nonincreasing in $\theta_t$. Hence, it suffices to impose the IR constraint for the highest-cost type:
\[
u_1(\bar{\theta})
+
\hat u_t(h_{t-1},\bar{\theta})
\geq0
\qquad
\text{for every }t\in\mathcal N
\text{ and }h_{t-1}\in\Theta^{t-1}.
\]

For fixed action rules, changing $u_1(\bar{\theta})$ shifts all continuation utilities by the same amount without affecting incentive compatibility. It also shifts every terminal payment by the same amount. The principal therefore chooses the smallest value satisfying all periodic IR constraints. Thus,
\[
u_1^*(\bar{\theta})
=
\inf
\left\{
u:
u+\hat u_t(h_{t-1},\bar{\theta})\geq0,\ 
\forall\,t\in\mathcal N,\ 
h_{t-1}\in\Theta^{t-1}
\right\}
=
-\inf_{\substack{
t\in\mathcal N\\
h_{t-1}\in\Theta^{t-1}
}}
\hat u_t(h_{t-1},\bar{\theta}).
\]

\end{proof}

 \subsection{Proof of Proposition \ref{prop:revenueequivalence}}

\begin{proof}
In the final period, $u_N(h_N) = p_N(h_N)-\theta_Nx_N(h_N)$. By Lemma~\ref{lem:utexpression}, $u_N(h_N) = u_1^*(\bar{\theta})+\hat u_N(h_N)$. Therefore, $p_N(h_N)
=
\theta_Nx_N(h_N)
+
u_1^*(\bar{\theta})
+
\hat u_N(h_N).$
Substituting the definition of $\hat u_N(h_N)$ yields the payment formula in the proposition.

By Lemma~\ref{lem:u1expression},
$u_1^*(\bar{\theta})$ is the smallest value satisfying all periodic
IR constraints. For fixed action rules, any increase in
$u_1(\bar{\theta})$ raises the terminal payment by the same amount at
every history. Hence, the displayed payment rule is the unique
implementable terminal payment rule that maximizes the principal's
expected payoff.
\end{proof}

\subsection{Proof of Proposition \ref{prop:summarystats}}

\begin{proof}
Fix a feasible value $u$ of $u_1(\bar{\theta})$, and define $A(c)\equiv\int_{\underline{\theta}}^c F(\theta)\,d\theta$. Let $B_1=0$, and, for $t\geq2$, define the payment balance inherited at the beginning of period $t$ by

\[
B_t(h_{t-1})
\equiv
\sum_{i=1}^{t-1}x_i c_i(h_{i-1})
-
\sum_{i=2}^{t-1}A\bigl(c_i(h_{i-1})\bigr).
\]

Let $W_t(B)$ denote the principal's optimal continuation payoff from
period $t$ onward when the inherited balance is $B$. By
Proposition~\ref{prop:revenueequivalence},
$p_N(h_N)=u+B_{N+1}(h_N)$. Hence, at the terminal node following
period $N$, the principal's payoff is $-u-B$, so define $W_{N+1}(B)=-u-B$.

By Lemma~\ref{lem:u1expression}, the period-$t$ IR constraint for the highest-cost type is $u+B_t(h_{t-1})-A\bigl(c_t(h_{t-1})\bigr)\geq0$ for $t\geq2$. Therefore, For $t=2,\ldots,N$, the continuation problem satisfies
\[
W_t(B) = \max_{\substack{c\in\Theta\\u+B-A(c)\geq0}} \left\{ F(c)\left[\alpha+W_{t+1}\bigl(B+c-A(c)\bigr)\right]
+ \bigl(1-F(c)\bigr)W_{t+1}\bigl(B-A(c)\bigr) \right\}.
\]
Indeed, work occurs with probability $F(c)$, in which case the next balance is $B+c-A(c)$, while shirk occurs with probability $1-F(c)$, in which case the next balance is $B-A(c)$. Since future costs are i.i.d., this problem depends on the past cost history only through $B$. Hence, an optimal cutoff can be selected in the form $c_t=\gamma_t(B_t)$.

It remains to show that $B_t$ is determined by the working-status history. The period-$1$ cutoff is independent of history, and $B_2=x_1c_1$, so $B_2$ is determined by $w_1$. For $t\geq2$, the balance evolves according to $B_{t+1} = B_t+x_t\gamma_t(B_t)-A\bigl(\gamma_t(B_t)\bigr)$.
Suppose inductively that $B_t=b_t(w_{t-1})$ for some function $b_t$. Then
\[
B_{t+1}
=
b_t(w_{t-1})
+
x_t\gamma_t\bigl(b_t(w_{t-1})\bigr)
-
A\!\left(\gamma_t\bigl(b_t(w_{t-1})\bigr)\right),
\]
which depends only on $w_t=(w_{t-1},x_t)$. Thus, by induction, $B_t=b_t(w_{t-1})$ for every $t$, and therefore
\[
c_t(h_{t-1})
=
\gamma_t\bigl(B_t(h_{t-1})\bigr)
=
\gamma_t\bigl(b_t(w_{t-1})\bigr)
\equiv
c_t(w_{t-1}).
\]
Since the argument applies to every feasible value of $u_1(\bar{\theta})$, optimizing over this value establishes that, without loss of optimality, the cutoff depends on the cost history only through the working-status history.
\end{proof}

    \subsection{Lemma \ref{lem:thresholdslowerbound} and the proof}

    \begin{lemma} \label{lem:thresholdslowerbound}
In any optimal mechanism, every cutoff at an on-path history is strictly greater than $\underline{\theta}$.
\end{lemma}

\begin{proof}
Fix an arbitrary optimal mechanism $M^*$ with cutoff rule
$\{c_t^*(w_{t-1})\}_{t=1}^{N}$. Suppose, toward a contradiction, that $c_t^*(w_{t-1})=\underline{\theta}$ at some on-path history $w_{t-1}$. Let
$q\equiv\Pr(w_{t-1})>0$. For sufficiently small $\varepsilon>0$, construct a mechanism $M^\varepsilon$ by setting $c_t^\varepsilon(w_{t-1}) = \underline{\theta}+\varepsilon$ and copying the continuation rule following shirk to the newly reached branch following work:
\[
c_k^\varepsilon(w_{t-1},1,v)
=
c_k^*(w_{t-1},0,v),
\qquad
k>t,\quad
v\in\{0,1\}^{k-t-1}.
\]
All other cutoffs remain unchanged, and the payment rule is adjusted according to Proposition~\ref{prop:revenueequivalence}. Let
\[
F_\varepsilon
\equiv
F(\underline{\theta}+\varepsilon),
\qquad
A_\varepsilon
\equiv
\int_{\underline{\theta}}^{\underline{\theta}+\varepsilon}
F(\theta)\,d\theta
\]
when $t\geq2$, and set $A_\varepsilon=0$ when $t=1$. By Lemma~\ref{lem:u1expression}, the resulting increase in $u_1^*(\bar{\theta})$ is at most $A_\varepsilon$. Indeed, the relevant $\hat u$-terms following shirk decrease by $A_\varepsilon$, while the corresponding terms following work are higher by
$\underline{\theta}+\varepsilon$; all other terms are unchanged. 

Apart from the possible increase in $u_1^*(\bar{\theta})$, the perturbation raises the principal's expected payoff by $q\left[ F_\varepsilon \bigl(\alpha-\underline{\theta}-\varepsilon\bigr) +A_\varepsilon \right]$.
Hence, the total payoff change satisfies
\begin{align*}
\Delta_\varepsilon
&\geq
qF_\varepsilon
\bigl(\alpha-\underline{\theta}-\varepsilon\bigr)
-
(1-q)A_\varepsilon \geq
qF_\varepsilon
\left(
\alpha-\underline{\theta}
-\frac{\varepsilon}{q}
\right),
\end{align*}
where the second inequality follows from
$A_\varepsilon\leq\varepsilon F_\varepsilon$. Since
$\alpha\geq\bar{\theta}>\underline{\theta}$, the last expression is strictly positive for sufficiently small $\varepsilon>0$. This contradicts the optimality of $M^*$. Therefore, $c_t^*(w_{t-1})>\underline{\theta}$ at every on-path history.
\end{proof}

    \subsection{Proof of Lemma \ref{lem:optmechproperties}}

    \begin{proof}

By Propositions~\ref{prop:revenueequivalence} and
\ref{prop:summarystats}, consider the full-tree version of the
objective in~\eqref{equ:optimprob1} on the entire product set
\(\Theta^{2^N-1}\), in which the minimum defining
\(u_1^*(\bar{\theta})\) is taken over all binary working-status
histories, including auxiliary histories. This full-tree objective is
continuous, since it is constructed from finitely many continuous
functions and a finite minimum. For any cutoff vector, its full-tree
value is weakly below the payoff of the induced mechanism, because
including auxiliary histories can only lower the minimum and thereby
raise \(u_1^*(\bar{\theta})\). Conversely, every mechanism admits the
canonical completion described in footnote~\ref{fn:auxiliary}, under
which auxiliary histories do not affect the minimum. Hence, the
full-tree problem and the original problem have the same supremum
value. Since \(\Theta^{2^N-1}\) is compact, the full-tree problem
admits a maximizer by the Weierstrass theorem, and the induced
mechanism of such a maximizer attains the common value. Therefore, an
optimal mechanism exists.

Fix an arbitrary optimal mechanism $M^*$, and let $\{c_t^*(w_{t-1})\}_{t=1}^{N}$ denote its cutoff rule. By the envelope formula, the agent's ex ante expected utility under $M^*$ is
\[
U^*
=
E_{\theta_1}\!\left[u_1(\theta_1)\right]
=
u_1^*(\bar{\theta})+\int_{\underline{\theta}}^{c_1^*} F(\theta) \, d\theta.
\]
Let $S^*$ denote the expected total surplus under $M^*$,
\[
S^* \equiv E_{\theta}\!\left[ \sum_{t=1}^{N}(\alpha-\theta_t)x_t^* \right]
 \leq N\bigl[\alpha-E(\theta)\bigr].
\]
Then, the principal's expected payoff is $V^*=S^*-U^*$.
The always-working mechanism is feasible and gives the principal the payoff $V^{\mathrm{aw}}
=
N\alpha-\left[\bar{\theta}+(N-1)E(\theta)\right]
=
N\bigl[\alpha-E(\theta)\bigr]
-
\bigl[\bar{\theta}-E(\theta)\bigr].$

Suppose that at least one initiation cutoff equals $\bar{\theta}$, and let $m\equiv\min\left\{ t\in\mathcal N: c_t^*(\mathbf{0})=\bar{\theta} \right\}$. First, suppose that $m\geq2$. By the definition of $m$, the history $\mathbf{0}_{m-1}$ occurs with positive probability. The period-$m$ IR constraint for the highest-cost type at this history requires
\[
0
\leq
u_m(\mathbf{0}_{m-1},\bar{\theta})
=
u_1^*(\bar{\theta})
-
\sum_{i=2}^{m} \int_{\underline{\theta}}^{c_i^*(\mathbf{0})} F(\theta) \, d\theta.
\]
Consequently,
\[
u_1^*(\bar{\theta})
\geq
\sum_{i=2}^{m} \int_{\underline{\theta}}^{c_i^*(\mathbf{0})} F(\theta) \, d\theta
\geq
\int_{\underline{\theta}}^{c_m^*(\mathbf{0})} F(\theta) \, d\theta
=\int_{\underline{\theta}}^{\bar{\theta}} F(\theta) \, d\theta
=
\bar{\theta}-E(\theta).
\]
Moreover, Lemma~\ref{lem:thresholdslowerbound} implies that $c_1^*>\underline{\theta}$, and hence $\int_{\underline{\theta}}^{c_1^*} F(\theta) \, d\theta>0$. Therefore,
\[
U^*
=
u_1^*(\bar{\theta})+\int_{\underline{\theta}}^{c_1^*} F(\theta) \, d\theta
>
\bar{\theta}-E(\theta).
\]
It follows that $V^*=S^*-U^*<N\bigl[\alpha-E(\theta)\bigr]-\bigl[\bar{\theta}-E(\theta)\bigr]=V^{\mathrm{aw}},
$ contradicting the optimality of $M^*$.

It remains to consider $m=1$, so that $c_1^*=\bar{\theta}$. The first-period IR constraint implies $u_1^*(\bar{\theta})\geq0$, and therefore $U^*
=
u_1^*(\bar{\theta})+\int_{\underline{\theta}}^{\bar{\theta}} F(\theta) \, d\theta
\geq
\bar{\theta}-E(\theta).$
Thus, 
\[
V^* \leq N\bigl[\alpha-E(\theta)\bigr] - \bigl[\bar{\theta}-E(\theta)\bigr] = V^{\mathrm{aw}}.
\]
Since $M^*$ is optimal and the always-working mechanism is feasible, equality must hold. In particular, $S^*=N\bigl[\alpha-E(\theta)\bigr]$. Because $f(\theta)>0$ on $\Theta$ and $\alpha-\theta>0$ for every $\theta<\bar{\theta}$, any on-path cutoff strictly below $\bar{\theta}$ would generate a positive probability of inefficient shirking and make the preceding inequality strict. Hence, every on-path cutoff equals $\bar{\theta}$, and $M^*$ induces the agent to work in every period.

Finally, suppose that no initiation cutoff equals $\bar{\theta}$. Then the all-shirking history $\mathbf{0}_{t-1}$ occurs with positive probability before every period $t$, so every initiation cutoff is on path. Lemma~\ref{lem:thresholdslowerbound} therefore implies $c_t^*(\mathbf{0}) \in (\underline{\theta},\bar{\theta})$ for every $t\in\mathcal N$. Thus, every optimal action rule either induces the agent to work in every period or has interior initiation cutoffs.
\end{proof}

\subsection{Proof of Proposition \ref{prop:constrainedoptim}}

\begin{proof}
Fix an optimal mechanism in the interior-initiation-cutoff case, with cutoff rule 
$\{c_t^*(w_{t-1})\}_{t=1}^{N}$. We first show that the IR constraint following the all-shirking history must bind. Suppose, to the contrary, that $-u_1^*(\bar{\theta})< -\sum_{i=2}^{N} \int_{\underline{\theta}}^{c_i^*(\mathbf{0})} F(\theta)\,d\theta$.
Because the inequality is strict and $c_N^*(\mathbf{0})<\bar{\theta}$, we can increase $c_N^*(\mathbf{0})$ by a sufficiently small $\varepsilon>0$ while preserving the inequality and keeping $c_N^*(\mathbf{0})+\varepsilon<\bar{\theta}$. Since the all-shirking term is the only term in the minimum defining $u_1^*(\bar{\theta})$ that depends on $c_N^*(\mathbf{0})$, this change leaves $u_1^*(\bar{\theta})$ unchanged.

Moreover, the probability of reaching $\mathbf{0}_{N-1}$ is strictly positive because all initiation cutoffs are interior. The resulting change in the principal's expected payoff is
\[
\Pr(\mathbf{0}_{N-1})
\int_{c_N^*(\mathbf{0})}^{c_N^*(\mathbf{0})+\varepsilon}
f(\theta)(\alpha-\theta)\,d\theta
>0,
\]
where the inequality follows from
$c_N^*(\mathbf{0})+\varepsilon<\bar{\theta}\leq\alpha$.
This contradicts optimality. Therefore,
\[
u_1^*(\bar{\theta})
=
-\widehat u_N(\mathbf{0}_N)
=
\sum_{i=2}^{N}
\int_{\underline{\theta}}^{c_i^*(\mathbf{0})}
F(\theta)\,d\theta.
\]
We may therefore replace $u_1^*(\bar{\theta})$ in the principal's
objective with $-\widehat u_N(\mathbf{0}_N)$ and impose the remaining
IR constraints. These constraints also ensure that
$\widehat u_N(\mathbf{0}_N)$ is indeed a minimizer in the definition of
$u_1^*(\bar{\theta})$. The binding argument above shows that no
optimum in the interior-initiation-cutoff case is excluded; making this substitution yields \eqref{equ:optimprob2}.

\end{proof}

\subsection{Proof of Theorem \ref{thm:optimalmechproperties}}

    \begin{proof}
Fix an arbitrary optimal mechanism $M^*$. By Lemma~\ref{lem:optmechproperties}, either $M^*$ induces the agent to work in every period, in which case it is an always-working mechanism, or its initiation cutoffs satisfy $c_t^*(\mathbf{0})\in(\underline{\theta},\bar{\theta})$ for every $t\in\mathcal N$. Consider the second case. By Proposition~\ref{prop:constrainedoptim}, the cutoff rule of $M^*$ maximizes the constrained problem in \eqref{equ:optimprob2}.

Holding the initiation cutoffs fixed, we show by backward induction that the objective in \eqref{equ:optimprob2} is maximized by setting every post-initiation cutoff equal to $\bar{\theta}$. Suppose that all post-initiation cutoffs after period $t$ have been set equal to $\bar{\theta}$; this is vacuous when $t=N$. Fix any history $w_{t-1}\neq\mathbf{0}_{t-1}$. Because the continuation rules following work and shirk in period $t$ are then identical, the part of the objective that depends on $c_t(w_{t-1})$ is
\[
\Pr(w_{t-1})
\left[
F(c_t(w_{t-1}))
\bigl(\alpha-c_t(w_{t-1})\bigr)
+
\int_{\underline{\theta}}^{c_t(w_{t-1})}
F(\theta)\,d\theta
\right]
\]
plus terms independent of $c_t(w_{t-1})$. Its derivative is $\Pr(w_{t-1})f(c_t(w_{t-1})) \bigl[\alpha-c_t(w_{t-1})\bigr] \geq0$,
where the inequality follows from $\alpha\geq\bar{\theta}$. It is strict at every on-path history whenever $c_t(w_{t-1})<\bar{\theta}$. Thus, proceeding backward from period $N$, the objective is weakly increased by setting
\[
c_t(w_{t-1})=\bar{\theta}
\qquad
\text{for every }t\geq2
\text{ and }w_{t-1}\neq\mathbf{0}_{t-1}.
\]
Every reachable post-initiation cutoff is set equal to $\bar{\theta}$; cutoffs at auxiliary histories are normalized to $\underline{\theta}$.

It remains to verify that the resulting cutoff rule is feasible. The initiation cutoffs are unchanged, so all IR constraints along the consistent-shirking histories remain satisfied. As established in Section~4.3, the remaining on-path IR constraints reduce to
\[
c_{t+1}^*(\mathbf{0})
+
\sum_{i=t+2}^{N}
\int_{\underline{\theta}}^{c_i^*(\mathbf{0})}
F(\theta)\,d\theta
+
E(\theta)-\bar{\theta}
\geq0,
\qquad
0\leq t\leq N-2.
\]
Since all initiation cutoffs are interior and the integral terms are nonnegative, the left-hand side is strictly greater than $\underline{\theta}+E(\theta)-\bar{\theta}$,
which is nonnegative by Assumption~\ref{assumption1}. Hence, the modified cutoff rule satisfies all the omitted IR constraints and is feasible for \eqref{equ:optimprob2}.

The modification therefore yields a feasible mechanism whose payoff is weakly greater than that of $M^*$. Since $M^*$ is optimal, the modified mechanism is also optimal. Moreover, if $M^*$ had any on-path post-initiation cutoff strictly below $\bar{\theta}$, the payoff improvement would be strict, contradicting optimality. Thus, up to the normalization of off-path cutoffs, every optimal action rule in the interior-initiation-cutoff case satisfies
\[
c_t(w_{t-1})=\bar{\theta}
\qquad
\text{whenever }w_{t-1}\neq\mathbf{0}_{t-1}.
\]
Together with the interior initiation cutoffs and the corresponding terminal payment rule, this is a consecutive-working menu. Therefore, under Assumption~\ref{assumption1}, an optimal mechanism is either a consecutive-working menu or an always-working mechanism.
\end{proof}

\subsection{Proof of Theorem \ref{thm:mainresult}}

\begin{proof}
Let $D(\alpha)\equiv V^*(\alpha)-V^{\mathrm{aw}}(\alpha)$. Because the always-working mechanism is feasible, $D(\alpha)\geq0$ for every $\alpha\geq\bar{\theta}$. Moreover, $V^{\mathrm{aw}}(\alpha)=N\alpha-\left[\bar{\theta}+(N-1)E(\theta)\right]$.

We first show that \(D\) is continuous and nonincreasing. For any
mechanism \(M\), let \(T_M\in[0,N]\) denote the expected total number
of periods in which the agent works. For any
\(\alpha_2>\alpha_1\), $V_M(\alpha_2)= V_M(\alpha_1)+(\alpha_2-\alpha_1)T_M$.
Let \(M_i\) be an optimal mechanism at \(\alpha_i\), \(i=1,2\).
Then $V^*(\alpha_2)
\geq
V_{M_1}(\alpha_2)
=
V^*(\alpha_1)+(\alpha_2-\alpha_1)T_{M_1}
\geq
V^*(\alpha_1)$,
whereas $V^*(\alpha_2)
=
V_{M_2}(\alpha_2)
=
V_{M_2}(\alpha_1)+(\alpha_2-\alpha_1)T_{M_2}
\leq
V^*(\alpha_1)+N(\alpha_2-\alpha_1)$.
Therefore,
\[
0
\leq
V^*(\alpha_2)-V^*(\alpha_1)
\leq
N(\alpha_2-\alpha_1),
\]
so \(V^*\) is $N$-Lipschitz continuous; hence \(D\) is continuous. Moreover,
\[
\begin{aligned}
D(\alpha_2)-D(\alpha_1)
&=
V^*(\alpha_2)-V^*(\alpha_1)
 -
\bigl[
V^{\mathrm{aw}}(\alpha_2)
 -
V^{\mathrm{aw}}(\alpha_1)
\bigr] \\
&=
V^*(\alpha_2)-V^*(\alpha_1)
 -
N(\alpha_2-\alpha_1)
\leq 0.
\end{aligned}
\]
Thus, \(D\) is nonincreasing.  Once the always-working mechanism becomes optimal, it remains optimal for all larger values of $\alpha$.

We next show that the always-working mechanism is optimal for all
sufficiently large $\alpha$. Define
\[
\Gamma(c)
\equiv
c+\frac{F(c)}{f(c)},
\qquad
\overline{\Gamma}
\equiv
\max_{c\in\Theta}\Gamma(c).
\]
Because $f$ is continuous and strictly positive on the compact set
$\Theta$, $\overline{\Gamma}<\infty$.

Consider any consecutive-working menu $M$ and hold its continuation
cutoffs fixed. Its expected payoff satisfies $\frac{\partial V_M}{\partial c_1} =f(c_1)\left[Q_M-\Gamma(c_1)\right]$,
where
\[
Q_M=\alpha[N-T_M(0)]-(N-1)E(\theta)-u_1^*(\bar{\theta})+P_M(0).
\]
Because $T_M(0)\leq N-1$, $P_M(0)\geq0$, and
\begin{align*}
u_1^*(\bar{\theta})=\sum_{t=2}^{N}\int_{\underline{\theta}}^{c_t(\mathbf{0})}F(\theta)\,d\theta<\sum_{t=2}^{N}\int_{\underline{\theta}}^{\bar{\theta}}F(\theta)\,d\theta=(N-1)[\bar{\theta}-E(\theta)],
\end{align*}
we have $Q_M>\alpha-(N-1)\bar{\theta}$.
Therefore, whenever $\alpha >(N-1)\bar{\theta}+\overline{\Gamma}$, we have $Q_M > \overline{\Gamma} \geq \Gamma(c_1)$, and hence $\frac{\partial V_M}{\partial c_1}>0$ for every interior $c_1$. Under Assumption~\ref{assumption1}, increasing $c_1$ slightly while holding the other cutoffs fixed preserves feasibility. Consequently, no consecutive-working menu can be optimal. By Theorem~\ref{thm:optimalmechproperties}, the always-working mechanism is then optimal for all sufficiently large $\alpha$.

Define $\hat{\alpha}\equiv\inf\left\{\alpha: D(\alpha)=0\right\}$. Note that $\hat{\alpha} \geq \bar{\theta}$, as the always-working mechanism is strictly suboptimal
whenever $\alpha<\bar{\theta}$. This is because it induces inefficient work
in the final period that can be eliminated without changing the
agent's promised continuation utility.%
\footnote{Replace the final-period full-work rule with the cutoff
$c=\alpha$ and adjust the terminal payment
to preserve the agent's expected continuation utility. This leaves
all earlier IC and IR constraints unchanged and increases the
principal's payoff by
$\int_c^{\bar{\theta}}(s-\alpha)f(s)\,ds>0$.}

By continuity,
$D(\widehat{\alpha})=0$. Since $D$ is nonincreasing and nonnegative, we have $D(\alpha)> 0$ for  $\bar{\theta}\leq\alpha<\widehat{\alpha}$, and $D(\alpha)=0$ for $\alpha\geq\widehat{\alpha}$.
Theorem~\ref{thm:optimalmechproperties} then implies that a consecutive-working menu is optimal for $\bar{\theta}\leq\alpha<\hat{\alpha}$, while the always-working mechanism is optimal for $\alpha\geq\hat{\alpha}$.
\end{proof}

\subsection{Proof of Theorem \ref{thm:generaloptimalmech}}

\begin{proof}
Fix an arbitrary optimal pay-at-the-end mechanism $M^*$ and an
on-path  history $w_N$. By
Lemma~\ref{lem:thresholdslowerbound}, every cutoff along this history
is strictly greater than $\underline{\theta}$. It therefore suffices
to show that, once an on-path cutoff equals $\bar{\theta}$, all
subsequent on-path cutoffs also equal $\bar{\theta}$.

Suppose that, following an on-path history $w_{s-1}$, $c_s^*(w_{s-1})=\bar{\theta}$. Since every type is prescribed
to work in period $s$, the envelope formula implies $u_s^*(w_{s-1},\theta_s) = u_s^*(w_{s-1},\bar{\theta}) + \bar{\theta}-\theta_s$.
Define $R \equiv u_s^*(w_{s-1},\bar{\theta})+\bar{\theta}$; then we have  $u_s^*(w_{s-1},\theta_s)=R-\theta_s$. The period-$s$ IR constraint for the highest-cost type implies $R\geq\bar{\theta}$.

Consider the following modification after $w_{s-1}$. Prescribe the
agent to work in every period from $s$ through $N$, and make the
terminal payment $P \equiv R+(N-s)E(\theta)$ if the agent follows the recommendation; otherwise, terminate the
relationship and make no payment. Under this continuation, a
period-$s$ type $\theta_s$ obtains expected utility
\[
P-\theta_s-(N-s)E(\theta)
=
R-\theta_s
=
u_s^*(w_{s-1},\theta_s).
\]
Thus, the modification preserves every period-$s$ type's
continuation utility and hence does not affect incentives or
participation at $w_{s-1}$ or at any preceding history.

Moreover, in any later period $j>s$, after the agent has followed
the work recommendations, his continuation utility from working
when his current cost is $\theta_j$ is
\[
P-\theta_j-(N-j)E(\theta) = R-\theta_j+(j-s)E(\theta) >0
\]
where the inequality follows from
$R\geq\bar{\theta}\geq\theta_j$ and $E(\theta)>0$.
Hence, the modified continuation is implementable.

The modification preserves the agent's expected continuation utility at $w_{s-1}$ and weakly increases total surplus, since work is efficient. The principal therefore weakly benefits from the modification. If any subsequent on-path cutoff were strictly below $\bar{\theta}$, then the increase in the principal's payoff would be strict. This would contradict the optimality of $M^*$. Therefore, $c_i^*\bigl(w_N^{1:i-1}\bigr)=\bar{\theta}$ for every $i>s$.

For the given on-path history $w_N$, if no cutoff equals
$\bar{\theta}$, set $t(w_N)=N$. Otherwise, let $s$ be the first
period for which
$c_s^*(w_N^{1:s-1})=\bar{\theta}$ and set
$t(w_N)=s-1$. By
Lemma~\ref{lem:thresholdslowerbound}, all cutoffs before period $s$
are interior, while the preceding argument implies that all cutoffs
from period $s$ onward equal $\bar{\theta}$. This establishes the two-stage structure.

\end{proof}

\end{appendices}

\bibliographystyle{econ} 
\bibliography{references}  

\newpage 
    \setcounter{page}{1}
    \setcounter{section}{0}
    \numberwithin{definition}{section}
    \numberwithin{theorem}{section}
    \numberwithin{lemma}{section}
    
    \renewcommand{\thesection}{B}

        \renewcommand{\thesubsection}{B.\arabic{subsection}}

    \begin{center}
        { \LARGE \textbf{Online Appendix to ``Dynamic Reward Design''} }
        
        \vspace*{20pt}
        
        {\large Yijun Liu}
    \end{center}

    \section{Online Appendix}

This appendix contains empirical examples of consecutive reward programs and
proofs of the results in Section~\ref{sec:extensions}.

    \begin{appendices}

\subsection{Empirical examples of consecutive reward programs}

\begin{figure}[H]
    \centering
    \setlength{\abovecaptionskip}{5pt}
    \setlength{\belowcaptionskip}{0pt}

    \begin{subfigure}[t]{0.48\textwidth}
        \vspace{0pt}
        \centering
        \includegraphics[
            width=\linewidth,
            height=0.21\textheight,
            keepaspectratio
        ]{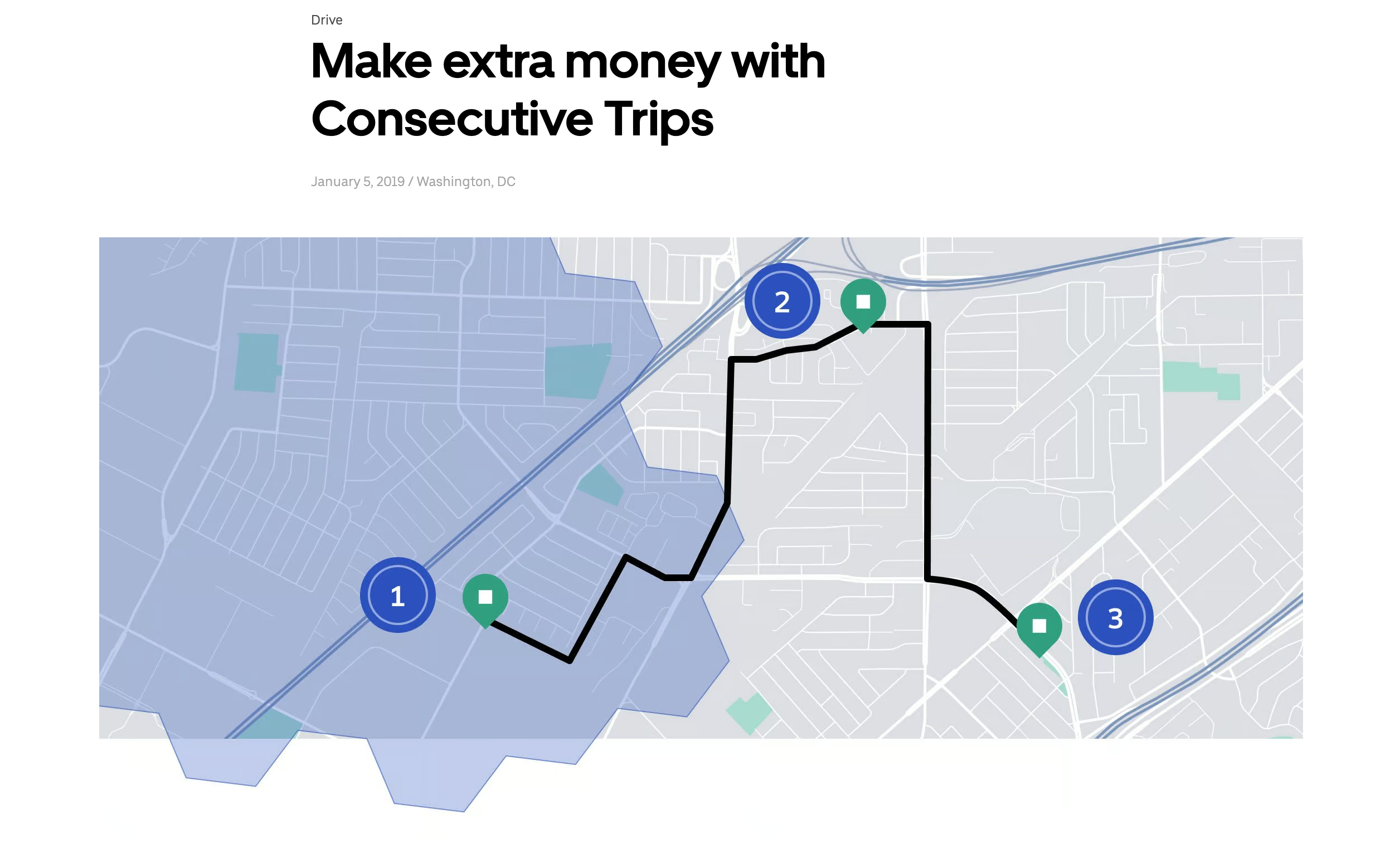}
        \caption{ \href{https://www.uber.com/us/en/blog/consecutive-trips-earnings/}{Uber}}
        \label{fig:uber}
    \end{subfigure}
    \hfill
    \begin{subfigure}[t]{0.48\textwidth}
        \vspace{0pt}
        \centering
        \includegraphics[
            width=\linewidth,
            height=0.21\textheight,
            keepaspectratio
        ]{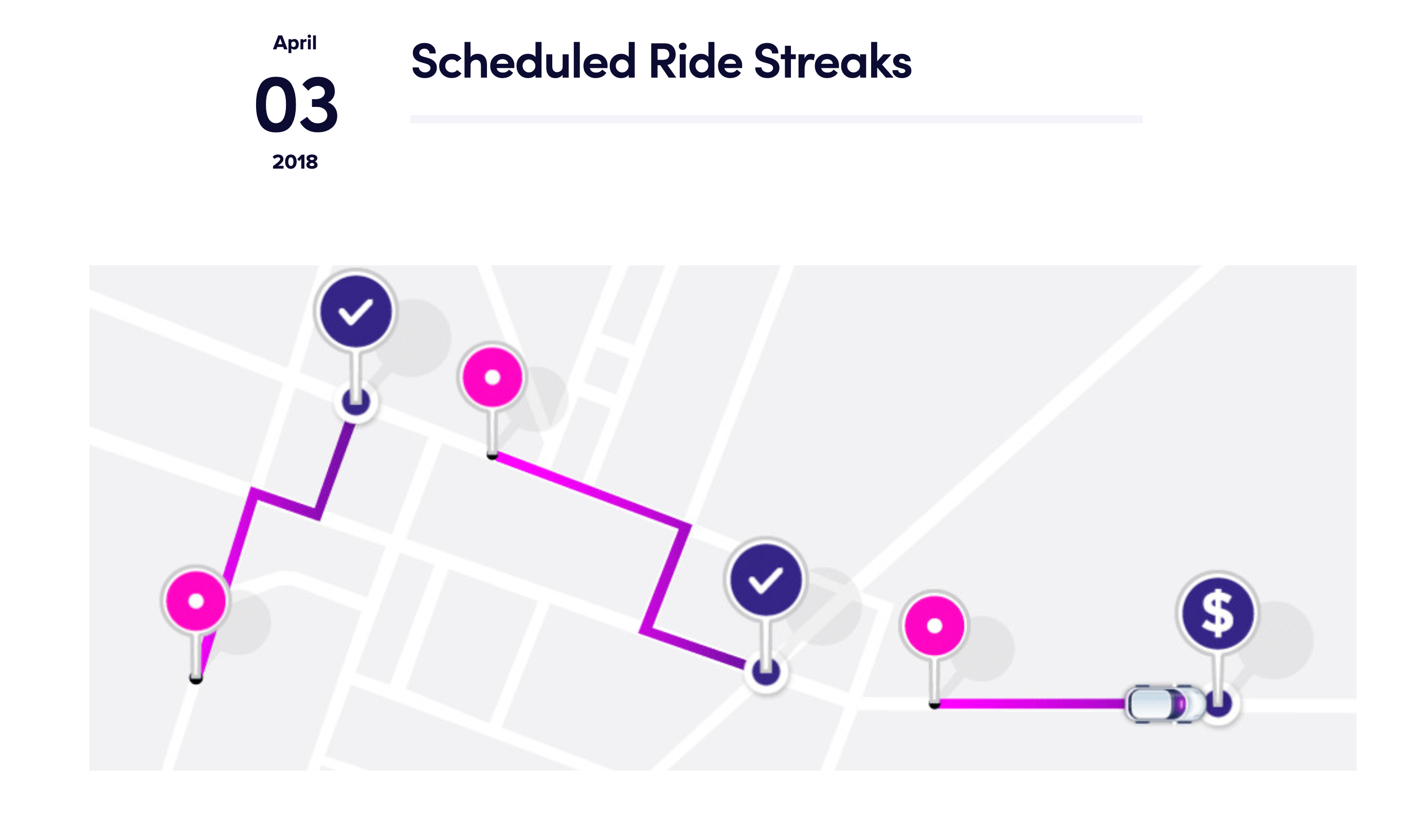}
        \caption{ \href{https://www.lyft.com/hub/posts/scheduled-ride-streaks}{Lyft} }
        \label{fig:lyft}
    \end{subfigure}

    \par\vspace{0.8em}

    \begin{subfigure}[t]{0.48\textwidth}
        \vspace{0pt}
        \centering
        \includegraphics[
            width=\linewidth,
            height=0.37\textheight,
            keepaspectratio
        ]{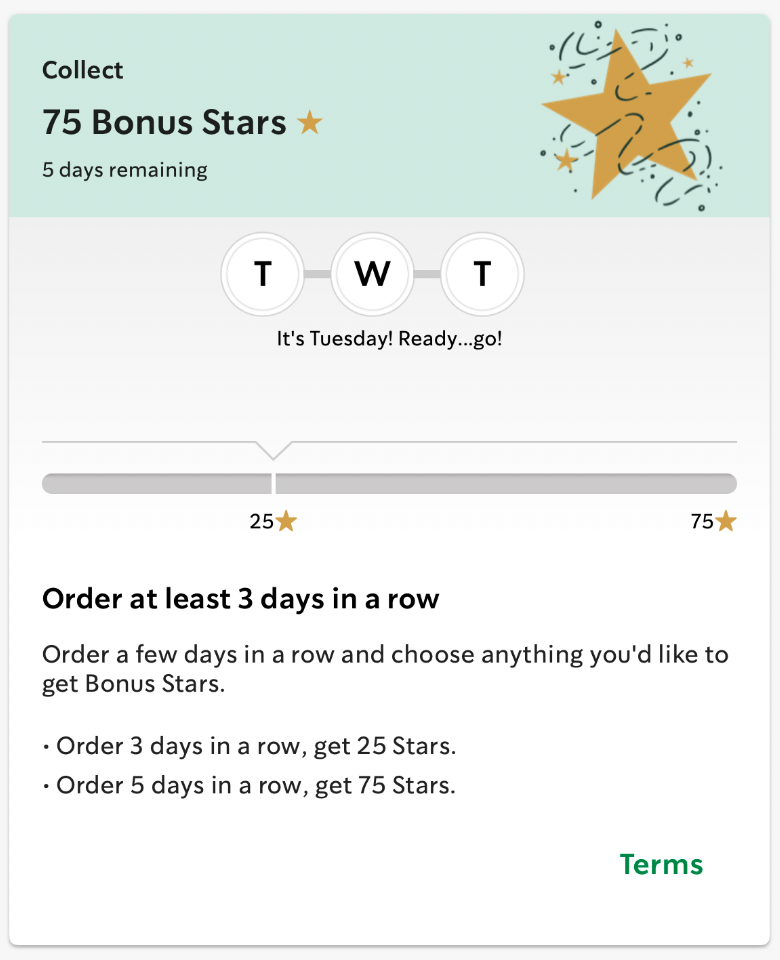}
        \caption{ \href{https://www.starbucks.com/rewards/terms/}{Starbucks} }
        \label{fig:starbucks}
    \end{subfigure}
    \hfill
    \begin{subfigure}[t]{0.48\textwidth}
        \vspace{0pt}
        \centering
        \includegraphics[
            width=\linewidth,
            height=0.37\textheight,
            keepaspectratio
        ]{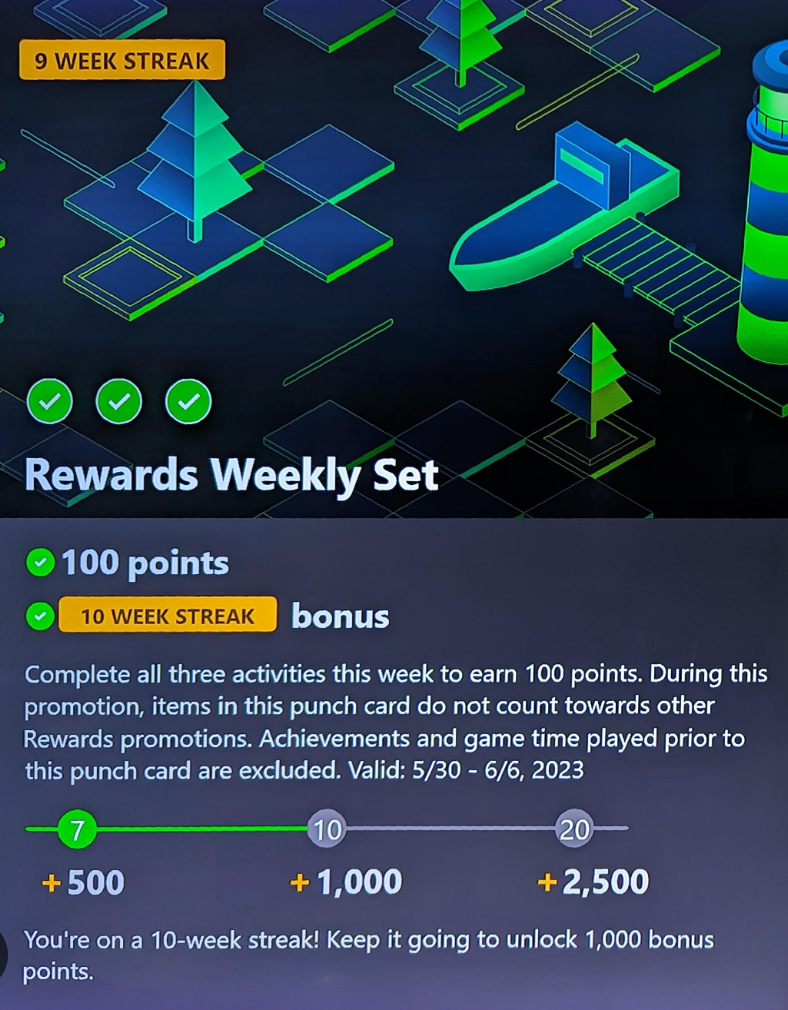}
        \caption{ \href{https://support.xbox.com/en-US/help/account-profile/manage-account/earn-points-microsoft-rewards}{Xbox} }
        \label{fig:xbox}
    \end{subfigure}
\end{figure}

\newpage

\begin{figure}[H]
    \ContinuedFloat
    \centering
    \setlength{\abovecaptionskip}{5pt}
    \setlength{\belowcaptionskip}{0pt}

    \setcounter{subfigure}{4}

    \begin{subfigure}[t]{0.55\textwidth}
        \vspace{0pt}
        \centering
        \includegraphics[
            width=\linewidth,
            height=0.40\textheight,
            keepaspectratio
        ]{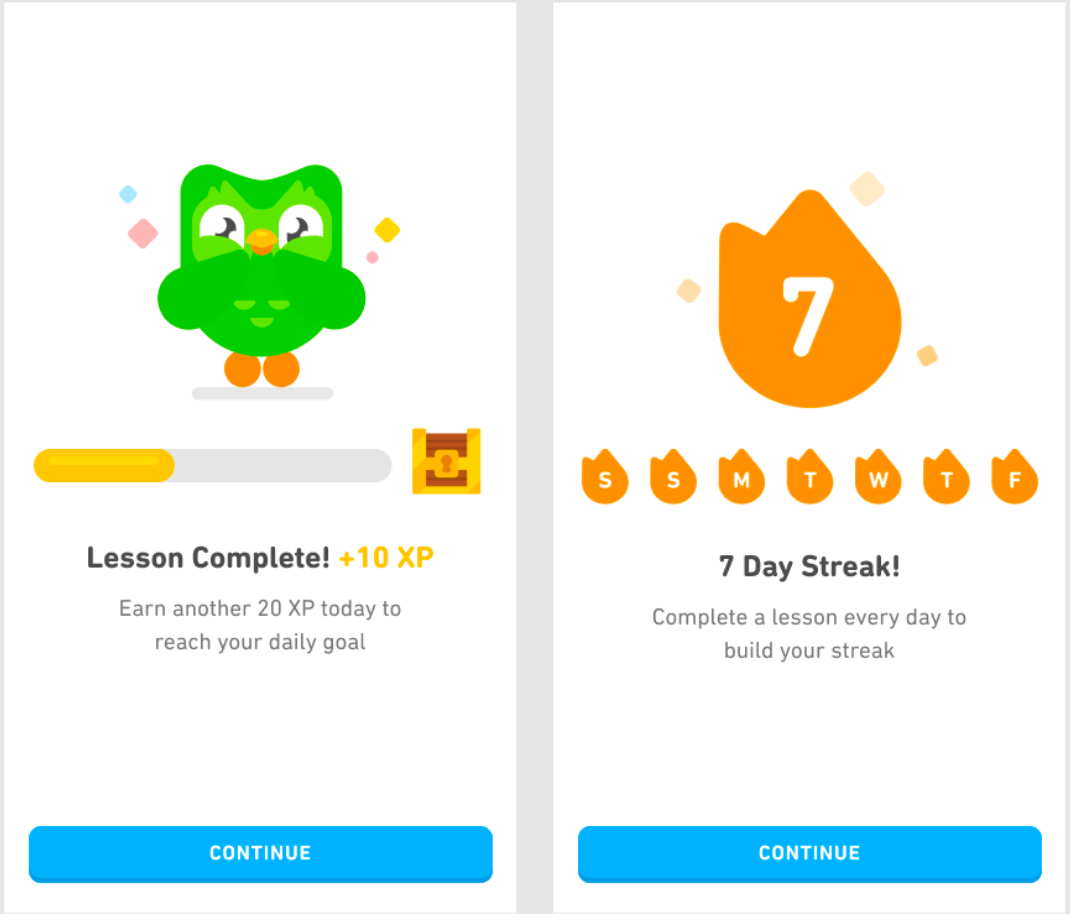}
        \caption{\href{https://www.duolingo.com/help/what-is-a-streak}{Duolingo} }
        \label{fig:duolingo}
    \end{subfigure}

    \caption{Empirical examples of consecutive reward programs.}
    \label{fig:empirical-reward-programs}

\end{figure}

\subsection{Proof of Proposition \ref{prop:stochasticimprov}}

\begin{proof}
Fix an optimal deterministic consecutive-working menu $M^*$, and let $c_1^*$ denote its first-period cutoff. Let $\pi(c)\equiv F(c)(\alpha-c)$ denote the principal's expected payoff in the static (one-period) problem, and let $x^{\mathrm{SB}}$ be the largest maximizer of $\pi$ over $\Theta$. We next show that $x^{\mathrm{SB}}<c_1^*$ and $\pi(x^{\mathrm{SB}})>\pi(c_1^*)$.

Let $\Delta V\equiv V_{M^*}(1)-V_{M^*}(0)$ denote the difference between the principal's continuation payoffs following work and shirk in period~1 under the consecutive-working menu $M^*$. The principal implements the first-best allocation in every subsequent period following work in period~1. Following shirk, however, every future initiation cutoff is interior, so some efficient work is omitted with positive probability. Hence, $\Delta V>0$.

Holding the continuation mechanisms fixed, the principal's payoff as a function of the first-period cutoff is 
\[
\Pi(c)=\pi(c)+F(c)\Delta V+V_{M^*}(0).
\]
Because $c_1^*$ is interior, its first-order condition gives $\pi'(c_1^*)=-f(c_1^*)\Delta V<0$, so $c_1^*$ is not a maximizer of $\pi$ and therefore $\pi(x^{\mathrm{SB}})>\pi(c_1^*)$.

Moreover, if $x^{\mathrm{SB}}>c_1^*$, then
\[
\Pi(x^{\mathrm{SB}})-\Pi(c_1^*)
=
\pi(x^{\mathrm{SB}})-\pi(c_1^*)
+
\Delta V\bigl[F(x^{\mathrm{SB}})-F(c_1^*)\bigr]
>0,
\]
contradicting the optimality of $c_1^*$. Equality is also impossible because $c_1^*$ is not a maximizer of $\pi$. Thus, $x^{\mathrm{SB}}<c_1^*$.

Let $u^*\equiv u_{1,M^*}^*(\bar{\theta})$ denote the $u^*_1(\bar{\theta})$ under $M^*$, and choose 
\[
0<\varepsilon< \min\left\{ 1,\, \frac{u^*}{\bar{\theta}-E(\theta)}\right\}.
\]
Consider the stochastic first-period action rule
\[
q_1^\varepsilon(\theta_1)
=
\begin{cases}
1,
& \theta_1<x^{\mathrm{SB}},\\
1-\varepsilon,
& x^{\mathrm{SB}}\leq\theta_1\leq c_1^*,\\
0,
& \theta_1>c_1^*.
\end{cases}
\]
Define the associated interim utility by $U^\varepsilon(\theta_1) \equiv u^* + \int_{\theta_1}^{\bar{\theta}} q_1^\varepsilon(s)\,ds$. Next, consider the following action rules for periods $t\geq2$: For $\theta_1>c_1^*$, retain the original mechanism. For $\theta_1\leq c_1^*$, prescribe work in every subsequent period following either realization of the first-period action. Conditional on following all recommendations, set the terminal payment equal to
\[
p_N^\varepsilon(\theta_1,x_1)
=
\begin{cases}
\theta_1+U^\varepsilon(\theta_1)+(N-1)E(\theta),
& \theta_1<x^{\mathrm{SB}},\ x_1=1,\\[1mm]
\theta_1+(N-1)E(\theta),
& x^{\mathrm{SB}}\leq\theta_1\leq c_1^*,\ x_1=1,\\[1mm]
\dfrac{U^\varepsilon(\theta_1)}{\varepsilon}
+(N-1)E(\theta),
& x^{\mathrm{SB}}\leq\theta_1\leq c_1^*,\ x_1=0.
\end{cases}
\]
If the agent disobeys a recommendation, the relationship is terminated and no payment is made.

We next verify implementability. For $\theta_1<x^{\mathrm{SB}}$, the agent works with probability one and obtains expected utility $U^\varepsilon(\theta_1)$. For
$\theta_1\in[x^{\mathrm{SB}},c_1^*]$, his expected utility is
\[
(1-\varepsilon)\cdot0
+
\varepsilon\cdot
\frac{U^\varepsilon(\theta_1)}{\varepsilon}
=
U^\varepsilon(\theta_1).
\]
For $\theta_1>c_1^*$, his utility remains $u^*$, which equals $U^\varepsilon(\theta_1)$ for $\theta_1$ in this range. Since $q_1^\varepsilon$ is nonincreasing and
\[
U^\varepsilon(\theta_1)
=
u^*
+
\int_{\theta_1}^{\bar{\theta}}
q_1^\varepsilon(s)\,ds,
\]
the first-period IC constraints are satisfied. The first-period IR constraints also hold because $U^\varepsilon(\theta_1)\geq u^*>0$.

For any $t\geq2$, following $x_1=1$, the agent's continuation utility from obeying is bounded below by $\theta_1+(t-1)E(\theta)-\theta_t \geq \underline{\theta}+E(\theta)-\bar{\theta} \geq0$, where the last inequality follows from Assumption~\ref{assumption1}. Following $x_1=0$, which occurs only for
$\theta_1\in[x^{\mathrm{SB}},c_1^*]$, his continuation utility is bounded below by
\[
\frac{U^\varepsilon(\theta_1)}{\varepsilon}
+
(t-1)E(\theta)-\theta_t
\geq
\frac{u^*}{\varepsilon}
+
E(\theta)-\bar{\theta}
>0.
\]
Thus, all subsequent IR constraints hold. Since future work is prescribed for every cost realization and the payment does not depend on future reports, all subsequent IC constraints are also satisfied. The payment rule is nonnegative, so the stochastic mechanism is implementable.

It remains to compare payoffs. Under $M^*$, every type
$\theta_1\leq c_1^*$ receives the terminal payment $c_1^*+u^*+(N-1)E(\theta)$. For types below $x^{\mathrm{SB}}$, the stochastic mechanism reduces this payment by
$\varepsilon(c_1^*-x^{\mathrm{SB}})$. For types in
$[x^{\mathrm{SB}},c_1^*]$, it reduces the expected payment by
$\varepsilon c_1^*$, while reducing the first-period work probability by $\varepsilon$. Therefore, the change in the principal's expected payoff is
\begin{align*}
\Delta^\varepsilon
={}&
\varepsilon F(x^{\mathrm{SB}})
\bigl(c_1^*-x^{\mathrm{SB}}\bigr)
+
\varepsilon
\bigl[F(c_1^*)-F(x^{\mathrm{SB}})\bigr]c_1^* -
\varepsilon\alpha
\bigl[F(c_1^*)-F(x^{\mathrm{SB}})\bigr]\\
={}&
\varepsilon
\left[
\pi(x^{\mathrm{SB}})
-
\pi(c_1^*)
\right]
>0.
\end{align*}
Hence, the stochastic mechanism strictly outperforms the optimal deterministic mechanism.
\end{proof}

\subsection{Proof of Proposition \ref{prop:stochasticimplementation}}

\begin{proof}
Fix $z_{t-1}$ and $\theta_t$, and suppress these arguments. Let $q\equiv q_t(z_{t-1};\theta_t)$ and $u\equiv u_t(z_{t-1};\theta_t)$, and denote the agent's conditional continuation utilities following work and shirk by $u^1\equiv u_t(z_{t-1};\theta_t,x_t=1)$ and $u^0\equiv u_t(z_{t-1};\theta_t,x_t=0)$.

By definition, $u=qu^1+(1-q)u^0$. If $q=0$, the work branch is off path, so its utility can be normalized to zero and the result is immediate. Hence, suppose that $q\in(0,1)$.

Following the work realization, prescribe work in every remaining period and make the terminal payment $P^1=\theta_t+(N-t)E(\theta)$ if the agent follows all subsequent recommendations. This gives the agent zero conditional utility in period $t$. Moreover, in any later period $j>t$, his continuation utility from working at cost $\theta_j$ is
\[
P^1-\theta_j-(N-j)E(\theta)
=
\theta_t-\theta_j+(j-t)E(\theta)
\geq
\underline{\theta}-\bar{\theta}+E(\theta)
\geq0,
\]
where the last inequality follows from Assumption~\ref{assumption1}. Thus, the first-best continuation following work is implementable with $u_t(z_{t-1};\theta_t,x_t=1)=0$.

The expected utility removed from the work branch is $qu^1$. Transfer this amount to the shirk branch by defining
\[
\delta\equiv\frac{qu^1}{1-q},
\qquad
\tilde u^0
\equiv
u^0+\delta
=
\frac{u}{1-q}.
\]
Then the agent's interim utility remains unchanged: $q\cdot0+(1-q)\tilde u^0 =qu^1+(1-q)u^0=u$.

If \(t=N\), there is no future continuation. Increase the terminal
payment following shirk by \(\delta\), so that the agent's conditional
utility on that branch becomes
\(\widetilde u^0=u^0+\delta\). Since
\((1-q)\delta=qu^1\), this exactly offsets the reduction in expected
payment from setting the work-branch utility to zero. Hence, the
agent's interim utility and the principal's expected payoff are
unchanged, and the result follows. For the remainder, suppose
\(t<N\).

We next show explicitly how the additional promise \(\delta\) can be
used to improve the continuation following shirk. Let \(C^0\) denote
the original continuation after shirk, with conditional utility
\(u^0\) and expected total surplus \(S^0\). Let
\(C^{\mathrm{fb}}\) denote the continuation that prescribes work in
every remaining period and pays
\[
P^0=\bar{\theta}+(N-t-1)\mathbb E(\widetilde{\theta})
\]
at the end if the agent follows all subsequent recommendations.This continuation is implementable and gives the agent conditional utility
\[
u^{\mathrm{fb}}=P^0-(N-t)E(\theta)=\bar{\theta}-E(\theta).
\]
Let $S^{\mathrm{fb}}$ denote its expected total surplus. Since work is efficient, $S^{\mathrm{fb}}\geq S^0$, with strict inequality whenever the original continuation after shirk fails to implement work for every cost realization.

If $\delta>0$, choose $\varepsilon>0$ sufficiently small that $\varepsilon \max\left\{ u^{\mathrm{fb}}-u^0,0 \right\} \leq\delta$. Following shirk, use $\mathcal C^{\mathrm{fb}}$ with probability $\varepsilon$ and $\mathcal C^0$ with probability $1-\varepsilon$. The resulting conditional utility before any additional transfer is $u^\varepsilon = (1-\varepsilon)u^0+\varepsilon u^{\mathrm{fb}} \leq u^0+\delta = \tilde u^0$.

Add the nonnegative amount $\tilde u^0-u^\varepsilon$ to every terminal payment on this branch. This preserves all future IC constraints, relaxes all future IR constraints, and gives the agent exactly the promised utility $\tilde u^0$. Moreover, because this construction is a public mixture of
implementable continuation mechanisms, it is itself implementable and
admits an outcome-equivalent reduced-form representation as a single
mechanism with stochastic action rules of the form defined above.

Under this lottery, a future work probability $q_j^0$ is replaced by $q_j^\varepsilon = (1-\varepsilon)q_j^0+\varepsilon = q_j^0+\varepsilon(1-q_j^0)$. Thus, the probability of work increases strictly at every future contingency at which the original continuation prescribed shirk with positive probability. Since the agent's conditional utility following shirk is held fixed at $\tilde u^0$, the principal's payoff gain on that branch is $\varepsilon\bigl(S^{\mathrm{fb}}-S^0\bigr)\geq0$, and is strictly positive whenever the original shirk continuation is not first-best.

Finally, the period-$t$ action rule and interim utility schedule are unchanged. Hence, the period-$t$ IC and IR constraints, as well as all constraints at preceding histories, remain satisfied. Since the agent's interim utility is unchanged and total surplus weakly increases, the principal's expected payoff weakly increases. Starting from an optimal continuation implementation, the modified implementation is therefore also optimal and satisfies the two properties stated in the proposition.
\end{proof}

\end{appendices}

\end{document}